\documentclass{ieeeaccess}

\usepackage{cite}
\usepackage{textcomp}
\usepackage{amsmath,amssymb,amsfonts}
\usepackage{pifont}
\usepackage{algorithm}
\usepackage{algorithmicx}
\usepackage{algpseudocode}
\usepackage[ansinew]{inputenc}
\usepackage[table,xcdraw]{xcolor}
\usepackage{mathtools}
\usepackage{graphicx}
\usepackage{caption}
\usepackage{subcaption}
\usepackage{import}
\usepackage{multirow}
\usepackage[export]{adjustbox}
\usepackage{breqn}
\usepackage{mathrsfs}
\usepackage{acronym}
\usepackage{setspace}
\usepackage{bm}
\usepackage{stackengine}
\usepackage{soul}
\usepackage{booktabs}
\usepackage{url}
\usepackage{listings}
\usepackage{hyperref}

\makeatletter

\newcommand{\cmark}{\ding{51}}%
\newcommand{\xmark}{\ding{55}}%

\AtBeginDocument{\DeclareMathVersion{bold}
\SetSymbolFont{operators}{bold}{T1}{times}{b}{n}
\SetSymbolFont{NewLetters}{bold}{T1}{times}{b}{it}
\SetMathAlphabet{\mathrm}{bold}{T1}{times}{b}{n}
\SetMathAlphabet{\mathit}{bold}{T1}{times}{b}{it}
\SetMathAlphabet{\mathbf}{bold}{T1}{times}{b}{n}
\SetMathAlphabet{\mathtt}{bold}{OT1}{pcr}{b}{n}
\SetSymbolFont{symbols}{bold}{OMS}{cmsy}{b}{n}
\renewcommand\boldmath{\@nomath\boldmath\mathversion{bold}}}
\renewcommand{\thetable}{\arabic{table}}

\makeatother

\def\BibTeX{{\rm B\kern-.05em{\sc i\kern-.025em b}\kern-.08em
    T\kern-.1667em\lower.7ex\hbox{E}\kern-.125emX}}

\begin{document}

\thispagestyle{empty}
\onecolumn
\begin{copyright}
  2024 IEEE. Personal use of this material is permitted. Permission from IEEE must be obtained for all other uses, in any current or future media, including reprinting/republishing this material for advertising or promotional purposes, creating new collective works, for resale or redistribution to servers or lists, or reuse of any copyrighted component of this work in other works.

  This document represents the accepted version of the article, revised by the author to address reviewer feedback and approved for publication by IEEE.
  \textbf{The final, published version is available on IEEE Xplore\textsuperscript{\tiny\textregistered} at the following \href{https://ieeexplore.ieee.org/document/10804604?}{link}}.
\end{copyright}

\clearpage
\pagenumbering{arabic}
\history{Date of publication 16 December 2024, date of current version 24 December 2024.}

\doi{10.1109/ACCESS.2024.3518483}

\title{Mobile Traffic Prediction at the Edge through Distributed and Deep Transfer Learning}

\author{\uppercase{Alfredo Petrella, Marco Miozzo\authorrefmark{1}, Paolo Dini\authorrefmark{1}}}

\address[1]{CTTC/CERCA, Av. Carl Friedrich Gauss, 7, 08860, Castelldefels, Barcelona, Spain (e-mails: petrellalfredo@gmail.com, marco.miozzo, paolo.dini@cttc.es}

\tfootnote{This publication has been partially funded by the Spanish project PID2020-113832RB-C22(ORIGIN)/MCIN/AEI/10.13039/50110001103 and the grant CHIST-ERA-20-SICT-004 (SONATA) by PCI2021-122043-2A/AEI/10.13039/501100011033}

\markboth
{Alfredo Petrella, Marco Miozzo, Paolo Dini: Mobile Traffic Prediction at the Edge through Distributed and Deep Transfer Learning}
{Alfredo Petrella, Marco Miozzo, Paolo Dini: Mobile Traffic Prediction at the Edge through Distributed and Deep Transfer Learning}

\corresp{Corresponding author: Alfredo Petrella (e-mail: petrellalfredo@gmail.com).}

\begin{abstract}
Traffic prediction represents one of the crucial tasks for smartly optimizing the mobile network. Recently, Artificial Intelligence (AI) has attracted attention to solve this problem thanks to its ability in cognizing the state of the mobile network and make intelligent decisions. Research on this topic has concentrated on making predictions in a centralized fashion, i.e., by collecting data from the different network elements and process them in a cloud center.
This translates into inefficiencies due to the large amount of data transmissions and computations required, leading to high energy consumption.

In this work, we investigate a fully decentralized AI solution for mobile traffic prediction that allows data to be kept locally, reducing energy consumption through collaboration among the base station sites.
To do so, we propose a novel prediction framework based on edge computing and Deep Transfer Learning (DTL) techniques, using datasets obtained at the edge through a large measurement campaign.
Two main Deep Learning architectures are designed based on Convolutional Neural Networks (CNNs) and Recurrent Neural Networks (RNNs) and tested under different training conditions.
Simulation results show that the CNN architectures outperform the RNNs in accuracy and consume less energy. In both scenarios, DTL contributes to an accuracy enhancement in 85\% of the examined cases compared to their \mbox{stand-alone} counterparts. Additionally, DTL significantly reduces computational complexity and energy consumption during training, resulting in a reduction of the energy footprint by 60\% for CNNs and 90\% for RNNs.
Finally, two \mbox{cutting-edge} eXplainable Artificial Intelligence techniques are employed to interpret the derived learning models.
\end{abstract}


\titlepgskip=-21pt

\maketitle

\section{Introduction}
\label{sec:introduction}

\PARstart{T}{raffic} analysis and prediction represent crucial tasks for mobile network operators to properly and efficiently manage the network, by implementing, e.g., network planning, Quality of Experience (QoE) management, anomaly detection algorithms, and cybersecurity frameworks. 
Historically, these tasks have been performed by mobile network operators through their data collected for billing, the Call Detail Records (CDRs). 
This solution relies on dedicated servers that retrieve the data from the base stations and then process it in the core part of the network. Therefore, data has to be transmitted from the edge to central servers, which implies possible privacy issues due to the transmission of sensitive information, higher network congestion probability, and higher energy consumption due to the massive volume of data to be exchanged. 

The availability of such a huge amount of data, together with the evolution in computational capabilities, allows taking advantage of Deep Learning (DL) to tackle multiple problems in network management. DL has been identified as one of the pillars of the sixth generation (6G) mobile networks to address intelligent management solutions \cite{Letaief2019}.
Nevertheless, there is a complexity downside that is typically not taken into account when envisioning these intelligent, autonomous, DL-based technologies.
In fact, Artificial Neural Networks (ANNs) adopted in DL are made up of many neurons and layers identified by a high number of trainable parameters to execute identified tasks with high accuracy.
This implies extremely high computational complexity, which requires an equally high energy consumption~\cite{DEVRIES20232191}.
From a mobile operator perspective, energy accounts for 20-40\% of its operational expenditures, pushing with urgency for cost reduction through network optimization due to the low revenue growth environment~\cite{gsma-green}. To do so, AI is expected to be the main pillar thanks to its ability in cognizing the state of the mobile network and make intelligent decisions~\cite{Luo2022}.
Therefore, it is of paramount importance to adopt sustainable design principles for AI~\cite{Schwartz2019} and advocate for efficiency as an evaluation criterion, alongside accuracy and other related measures. 

Multi-access Edge Computing (MEC) represents an efficient solution to address data privacy, network congestion, and energy sustainability issues~\cite{thembelihle2017softwarization}. The main idea behind MEC is to process data directly at the edge, without sending it to central servers that may be managed by a third party. This approach solves both privacy and congestion concerns. Moreover, it has been demonstrated that MEC can save up to $25\%$ of the network energy consumption~\cite{Ahvar2019}. These savings are mainly enabled by reduced communication costs (avoiding data transmission to the central server) and smaller and energy-efficient devices, which use less power for refrigeration compared to data centers. 

In line with the above, we propose a distributed DL solution for traffic prediction implemented directly at the base stations.
Deep Transfer Learning (DTL)~\cite{iman2023review} is leveraged to additionally reduce the complexity of the training phase and, consequently, the energy consumed. The main principle of DTL is to exploit the knowledge of a teacher network trained with general data and then retrain student networks on a more specific local dataset. This allows the reduction of the computational capability required by the edge devices~\cite{Sharma2018}. In particular, we propose to exploit the learned knowledge for the other student base stations in a collaborative and distributed fashion.

In addition, we provide insights on the DL models built for traffic prediction through \mbox{Visual-based} eXplainable AI (\mbox{v-XAI}) \cite{XAI-survey}. These tools give explanations of \mbox{black-box} models to reveal their behavior and underlying \mbox{decision-making} mechanisms with a specific focus on how to visually represent their results for a general audience. The analysis performed allows for a better understanding of the output of both DL models and the DTL process.

In this work, data collected from several operative base stations located in Barcelona, Spain are used. Data is collected with OWL \cite{bui2016owl}, a tool capable of decoding the unencrypted  Physical Downlink Control Channel (PDCCH) of the Long Term Evolution (LTE) radio transmission technology. Among the control data contained in the PDCCH, the Downlink Control Information (DCI) messages can be retrieved, which contain \mbox{radio-link} level settings for the user communication, both for the downlink and uplink channels.
This information includes, among others, the Radio Network Temporary Identifier (RNTI) (a temporary ID assigned to each UE), the Modulation and Coding Scheme (MCS), and the number of allocated Resource Blocks (RBs) per frame.
Therefore, it is possible to infer the resources allocated to the users and the throughput of the base station without accessing the actual transmitted data, guaranteeing higher user privacy. The presented solution, based on control data and avoiding usual deep packet inspection techniques from the user plane, reduces the memory footprint for processing and includes an additional level of privacy since it does not need to access the user-generated content. 

Although the literature contains papers using classical centralized learning in data center for mobile traffic prediction, only few works have explored a distributed paradigm. In particular, a fully decentralized framework, where no central servers are required, is still lacking at the time of this writing, as described in Section~\ref{sec:related_work}.
We investigate Convolutional Neural Networks (CNNs) and Recurrent Neural Networks (RNNs) both with standard \mbox{stand-alone} databases and with the DTL paradigm. Simulation results show that the predictions in the local datasets present good levels of accuracy. The models evaluated with DTL have improved accuracy performance with respect to their corresponding stand-alone versions. All the results obtained are in line with benchmark solutions based on Support Vector Regression (SVR) \cite{liu2015empiricalSVR} applied to the \mbox{stand-alone} database. The analysis also details the difficulties in applying Transfer Learning (TL) to SVR. Moreover, we measure the complexity of the proposed models together with the energy savings obtained by the adoption of DTL, which can reach up to 60\% for CNN and 90\% for RNN. Finally, the learned models are interpreted through the XAI SmoothGrad \cite{SmoothGrad} and \mbox{Layer-wise} Relevance Propagation (LRP) \cite{LRP} algorithms. It is to be noted that the XAI analysis represents a novel contribution to this field, as it has never been applied to mobile traffic prediction models, to the best of our knowledge.

As a result, the original contributions of the paper are summarized in the following list.
\begin{itemize}
    \item We design a novel and efficient MEC-based framework for collecting and processing mobile data obtained by passively sniffing LTE control channel from different base stations. An Exploratory Data Analysis (EDA) for evaluating the main characteristics of the different datasets is also provided.
    \item Two mobile traffic prediction models based on RNNs and CNNs are designed. The performance of the proposed models is assessed with both \mbox{stand-alone} and DTL paradigms. A comparison with \mbox{SVR-based} solutions as a benchmark is also presented.
    \item The complexity and the energy footprint of the proposed models are analyzed.
    \item The interpretation of the obtained models is discussed by applying two XAI algorithms (SmoothGrad and LRP).
\end{itemize}

The rest of the paper is organized as follows. The background on MEC and the architectural scenario for traffic profiling at the edge is introduced in Section~\ref{sec:background}.
The previous literature and the innovative aspects of this work are presented in Section~\ref{sec:related_work}.
The datasets used and the analysis of their main features are described in Section~\ref{sec:datasets}. Section~\ref{sec:models} reports the description of the tackled tasks and the presented DL models and architectures.
Section~\ref{sec:results} contains a dissertation about the models' training, accuracy and complexity performance together with their energy consumption. 
In Section~\ref{sec:explainability}, the outcomes of the two XAI techniques are discussed. 
Finally, Section~\ref{sec:conclusion} concludes the paper with remarks and possible future directions.

\section{Background}
\label{sec:background}

Recently, the architectural availability of edge computing resources to execute AI directly at the edge has attracted significant attention. From the service side, it would help to support Ultra-Reliable Low Latency Communications (URLLC) scenarios, like factory automation, autonomous driving, remote surgery, and augmented/virtual reality~\cite{gupta20216g}. However, locating the data processing in the proximity of its source enables many more benefits:
\begin{itemize}
	\item \emph{Computation}: the algorithms process only local data, which implies smaller datasets and, thus, the use of less demanding hardware, both in terms of computational and memory requirements. 
	\item \emph{Communication}: proximity allows for the reduction of data transmission across different elements, thereby alleviating network congestion.
	\item \emph{Privacy}: distributed data avoids its passage through different network elements, which prevents leakage~\cite{Parikh2019}.
	\item \emph{Energy}: the advantages in computation and communication enable a reduction in energy used.
\end{itemize}

Regarding the possible applications, AI will play an important role in providing solutions to the resource management problem in mobile communications~\cite{Luo2022}. Typical examples are designing and optimizing 6G architectures, protocols, and operations.
In particular, the traffic load significantly affects the quality of experience (QoE) perceived by users, as their connections have to share the network resources. Based on this traffic load, an intelligent cognitive engine can help in reconfiguring the mobile network to avoid traffic congestion during high peaks, reducing energy consumption when traffic is low~\cite{Luo2022}.
However, predicting the traffic load is not trivial since it is correlated to many parameters, e.g., channel usage, link conditions, users' mobility patterns, etc.
In such cases, the challenges are on the model definition of complex 6G elements, which often has to be defined as a tractable Markov decision process, and the algorithm deployment, since it has to work online. Consequently, a trade-off between optimality and efficiency has to be found. In the case of Radio Resource Management (RRM), individual rule-based algorithms can be replaced by a general-purpose learning framework capable of autonomously generating complex algorithms specialized for each RRM functionality. 
To correctly perform RRM, traffic prediction plays a crucial role. By providing such functionality directly at the edge, the aforementioned benefits can be leveraged, leading to more accurate traffic prediction algorithms that consume fewer resources both from communication and energy perspectives.
In this respect, the current operational state of 5G is proving to be unsustainable due to the vast volume of data being generated, transmitted, processed, and stored~\cite{cheng20225g}. Moreover, data will experience exponential growth in the future, and by 2035, it is expected to reach an annual total of 2000 zettabytes, constituting approximately 20\% of the world's total energy consumption~\cite{statista}.
Thus, edge computing represents an important enabler toward an AI-native environmentally sustainable 6G~\cite{Mao2022}.

However, distributing the computation requires the use of specific techniques. The \emph{edge intelligence} paradigm, also called \emph{edge AI}~\cite{Zhou2019, Deng2019}, aims at evaluating distributed solutions to run ML models (the \emph{inference} phase) and to train ML models (the \emph{training} phase).
With respect to the training phase, the main problem of a distributed solution is the convergence of a consensus and how to synchronize and update the local gradients. The most popular solution for distributed training is represented by Federated Learning (FL)~\cite{Chen2019}. In this solution, the server is in charge of combining the results of the local models' training. However, the local learning methods are based on Stochastic Gradient Descent (SGD) and thus they are not optimized for working with unbalanced and non-Independent and Identical Distribution (IID) data. 

\begin{figure}[h!]
	\centering
	\includegraphics[width=\linewidth]{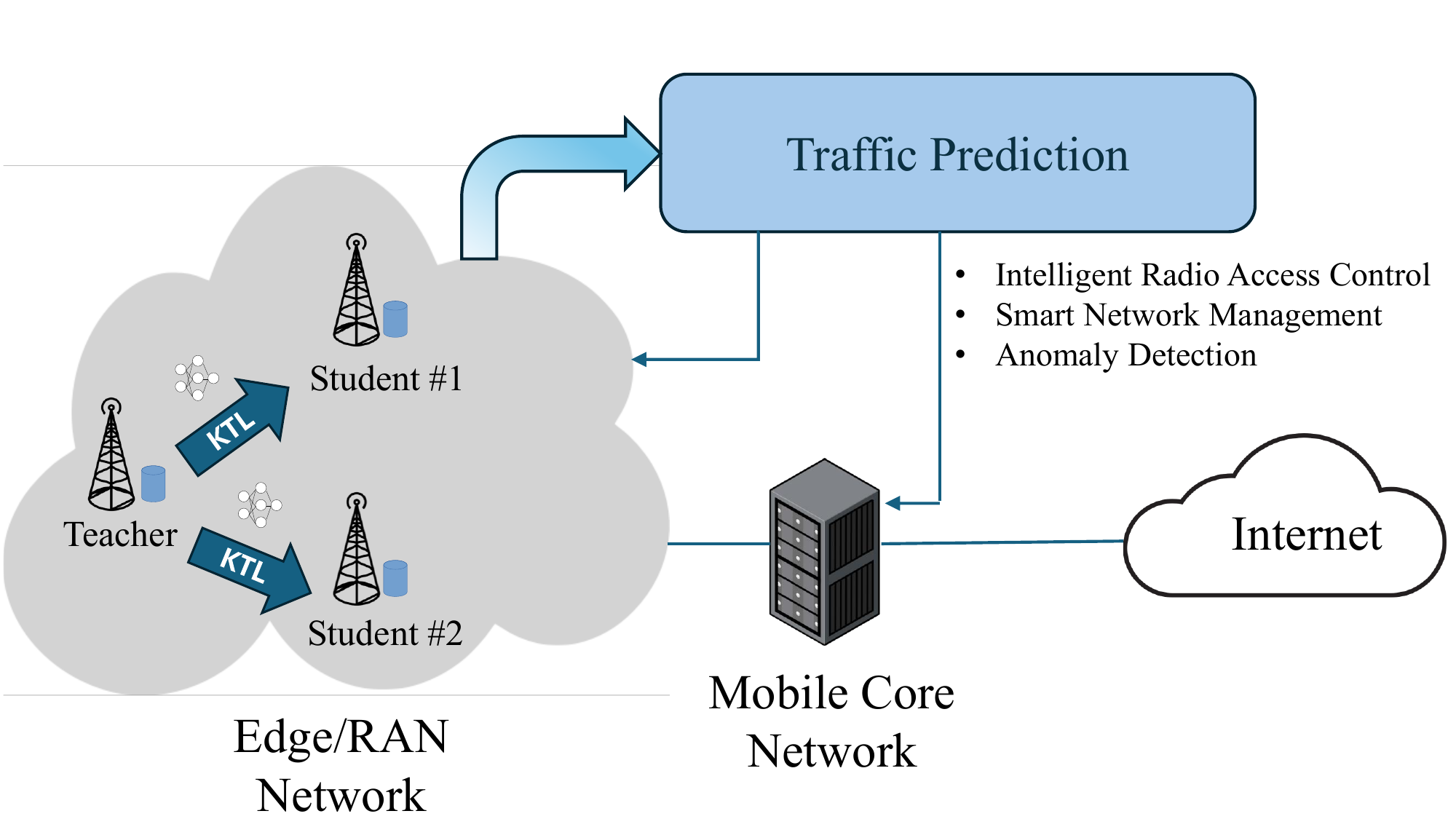}
	\caption{Example of the Edge AI scenario in mobile networks.}\label{fig:scenario}
\end{figure}

In this work, we are interested in evaluating collaborative solutions for training ML models for traffic prediction based on DTL. The aim is to avoid the synchronization problems of FL and be able to use the acquired knowledge from one BS site to others to save energy while maintaining similar accuracy with respect to benchmark centralized solutions. An example of the considered scenario is provided in Fig.~\ref{fig:scenario}.
Thanks to the proposed solution operators can learn the traffic patterns of their users in a specific area covered by a base station and thus optimize the management of such area, for example in terms of resource allocation (i.e., intelligent radio access control and smart network/device management~\cite{docomo}) and anomaly detection (i.e., identify different types of anomalous events generated by flash crowds~\cite{Pelati2022}). This can be done without moving data with privacy issues from the edge. Moreover, the DTL framework enables the usage of the knowledge acquired from one antenna site to be used in another site, thus saving energy thanks to the reduced computation and communication (i.e., only the ML model has to be exchanged among the sites). 
Furthermore, in this work data from the unencrypted control channel of an operative mobile network is exploited to properly characterize the mobile traffic patterns at the base station site, thus maintaining a higher level of privacy for users' data and reducing storage and monitoring processing, as detailed in Section~\ref{sec:related_work}.
\section{Related Work}
\label{sec:related_work}

Traffic prediction has been tackled for years due to its relevance for mobile network management. Traditional approaches involve SVR and AutoRegressive Integrated Moving Average (ARIMA) models \cite{yu2010networkARIMA} \cite{biernacki2017improvingARIMA}.
In the last few years, thanks to the evolution in computational capabilities, DL methods have overcome the traditional Machine Learning (ML) models. This is due to DL's structural ability to successfully recognize complex patterns and relationships present in mobile traffic time series \cite{bui2017survey}, and in other \mbox{networking-related} tasks \cite{surveyIntroPrevious1} \cite{surveyIntroPrevious2}. 
Most of the literature regarding traffic prediction with DL is characterized by a centralized approach.
In this category, good traffic prediction results have been obtained in \cite{zhang2018citywide} through densely connected \mbox{two-dimensional} CNNs, which are able to spot both the spatial and temporal dependence of cell traffic. In \cite{wang2017spatiotemporal}, an \mbox{autoencoder-based} deep model for spatial modeling and Long \mbox{Short-Term} Memory units (LSTMs) for temporal modeling was proposed and tested using a dataset provided by China Mobile. Similarly, in \cite{hua2019deep} LSTMs equipped with stochastic connectivity were proposed with the aim of reducing the considerable computing cost of the corresponding vanilla version.
Combinations of the two architectures mentioned above were effective as well, as presented in \cite{huang2017study} \cite{zhang2018long}.

However, in centralized approaches data needs to be transmitted from the edge to dedicated servers. To deal with such issues, distributed approaches based on FL have been recently studied, where only local ML models are exchanged with a central server.
In \cite{Zhang2022} CDR was used to train models for different neighborhoods of the city of Milan with a FL paradigm. In addition, Model-Agnostic Meta-Learning (MAML) has been used to train a sensitive initial model that can adapt to heterogeneous scenarios in different regions.
It is to be noted that, the majority of the cited works use CDR, i.e., mobile network operators' data which is used for billing purposes. This approach clearly implies potential privacy leaks, as well as a large amount of data to be exchanged within the network, resulting in considerable energy consumption.

To address the privacy issue, the same datasets considered in this work have already been adopted in~\cite{Perifanis2023}, where the authors present an FL-based solution to show its effectiveness considering the non-IID-ness of the data and different federated aggregators. Simulation results show that the proposed FL-based solutions reduce carbon emissions and energy consumption. However, the FL paradigm is based on a two-tier architecture where the MEC server has to coordinate during the learning phase with all the clients located in the proximity of the data sources. This might slow down the training process due to high network latency, unreliable links, or straggled clients. In addition, the central MEC server can be impacted by the single point of failure problem, i.e., if it becomes unreachable due to network issues or an attack, the training process is halted. Similarly, it may also become a bottleneck when the number of clients is very large.

Alternatively, this work aims to demonstrate the advantages of building one generic model to be trained directly on distributed sources of data (i.e., the base stations) to perform prediction and to apply DTL from the different sites.

Finally, to the best of our knowledge, XAI algorithms have never been applied to mobile traffic prediction models. Therefore, this work represents the first attempt to extract insights about the way models combine input variables to perform highly accurate predictions and to investigate how to improve model training when applying DTL. 
Similarly, the energy footprint assessment represents a unique contribution to this field. Table~\ref{table:related} provides a summary of the key points discussed in this section.

\begin{table*}[tb]%
  \caption{Comparison of Related Work approaches with the Proposed Solution.}
\begin{center}
\begin{tabular}{c|c|c|c|c|c|c}
\toprule
                              & Centralized     & \begin{tabular}[c]{@{}c@{}}Distributed with\\ central server\\ (federated)\end{tabular} & \begin{tabular}[c]{@{}c@{}}Distributed\\ server-less\\ (de-centralized)\end{tabular} & \begin{tabular}[c]{@{}c@{}}Privacy\\ protection\end{tabular}  & XAI-evaluated & \begin{tabular}[c]{@{}c@{}}Energy footprint\\ assessment\end{tabular}\\
\midrule
\hline
\cite{zhang2018citywide}      & \cmark          & \xmark        & \xmark        & \xmark        & \xmark        & \xmark        \\
\hline
\cite{wang2017spatiotemporal} & \cmark          & \xmark        & \xmark        & \xmark        & \xmark        & \xmark        \\
\hline
\cite{hua2019deep}            & \cmark          & \xmark        & \xmark        & \xmark        & \xmark        & \xmark        \\
\hline
\cite{huang2017study}         & \cmark          & \xmark        & \xmark        & \xmark        & \xmark        & \xmark        \\
\hline
\cite{zhang2018long}          & \cmark          & \xmark        & \xmark        & \xmark        & \xmark        & \xmark        \\
\hline
\cite{Zhang2022}              & \xmark          & \cmark        & \xmark        & \xmark        & \xmark        & \xmark        \\
\hline
\cite{Perifanis2023}          & \xmark          & \cmark        & \xmark        & \cmark        & \xmark        & \xmark        \\
\hline
Proposed Solution             & \xmark          & \xmark        & \cmark        & \cmark        & \cmark        & \cmark        \\
\hline
\bottomrule
\end{tabular}
\end{center}
\label{table:related}
\end{table*}
\section{Data exploration}
\label{sec:datasets}

The datasets have been obtained leveraging the information collected at the base stations' level through  OWL \cite{bui2016owl}. OWL is an \mbox{open-source} tool that allows decoding the DCI messages carried in the LTE PDCCH, which contains \mbox{radio-link} level settings for the user communication between the User Equipments (UEs) and the base station, both for the downlink and uplink channels. 
The data considered in this paper was collected from three different base stations located in the districts of Poble-sec (PS), El Born (EB), and Les Corts (LC), in the metropolitan area of Barcelona, Spain. The first one consists of approximately \mbox{twenty-eight} days of records, while the others have seven and twelve days of history, respectively.

The datasets are preliminary parsed, reorganized, and downsampled with a granularity of two minutes to mitigate the effect of missing values in the resulting dataset. In fact, the data collection phase is affected by decoding errors, resulting in missing values for the variables stored in the messages. 
Instances in which at least one variable is missing are removed, totaling roughly seventeen minutes from PS, six minutes from EB, and nine minutes from LC, which represent less than 2\textperthousand{} of the original datasets. Thus, downsampling allows a reduction in the number of instances where the value of at least one variable is missing for the entire time window and, in turn, enables the usage of DL schemes.

In this work, 5 variables are considered based on their importance with respect to the traffic prediction problem studied, namely:
\begin{itemize}
	\item \textit{RNTI}$_{count}$: the number of unique RNTIs; it can be interpreted as the number of users that communicate at least once with the base station within each time window. 
	\item \textit{MCS}$_{down}$ and \textit{MCS}$_{up}$: the average MCS assigned in downlink and uplink within each time window. 
	\item \textit{RB}$_{down}$ and \textit{RB}$_{up}$: the fraction of assigned RBs in downlink and uplink over the maximum number of available RBs within each time window; in the considered scenario this value ranges in $[0,100]\cap \mathbb{N}$, which corresponds to the case of 20 MHz of channel bandwidth.
\end{itemize}

\begin{figure}[h!]
	\centering
	\includegraphics[width=\linewidth]{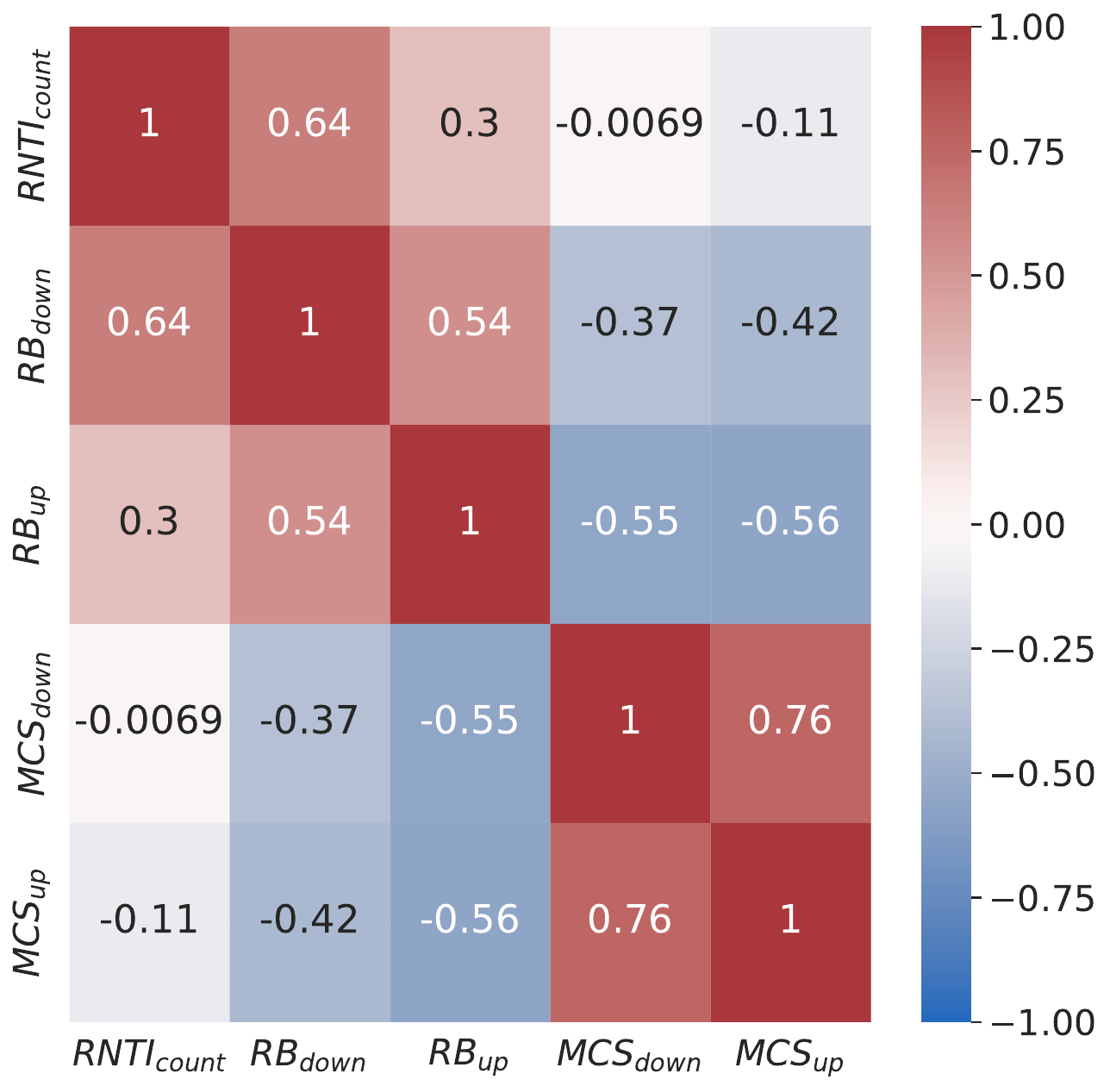}
	\caption{Average Pearson correlation matrix across the three datasets, each containing the same five variables. This matrix represents the average pairwise Pearson correlation coefficients calculated for the three datasets.}\label{fig:Pearson}
\end{figure}

The resulting correlation among the considered variables is reported in Fig.~\ref{fig:Pearson} using Pearson coefficient. The matrix shows a poor correlation among the 5 data features. In particular, \textit{RB}$_{down}$ is more correlated to \textit{RNTI}$_{count}$ than \textit{RB}$_{up}$. This is due to the fact that downlink traffic is predominant in LTE with respect to uplink. 
Finally, it is interesting to note that the MCS columns are negatively correlated with the percentage of assigned RBs. This is because a low average MCS, whether in the downlink or uplink, indicates degrading channel quality, requiring more RBs to transmit the same amount of data.

\begin{figure*}
	\centering
	\subfloat[\protect\label{subfig:PS_variables_rnti_count}]{
		\includegraphics[width=.33\textwidth]{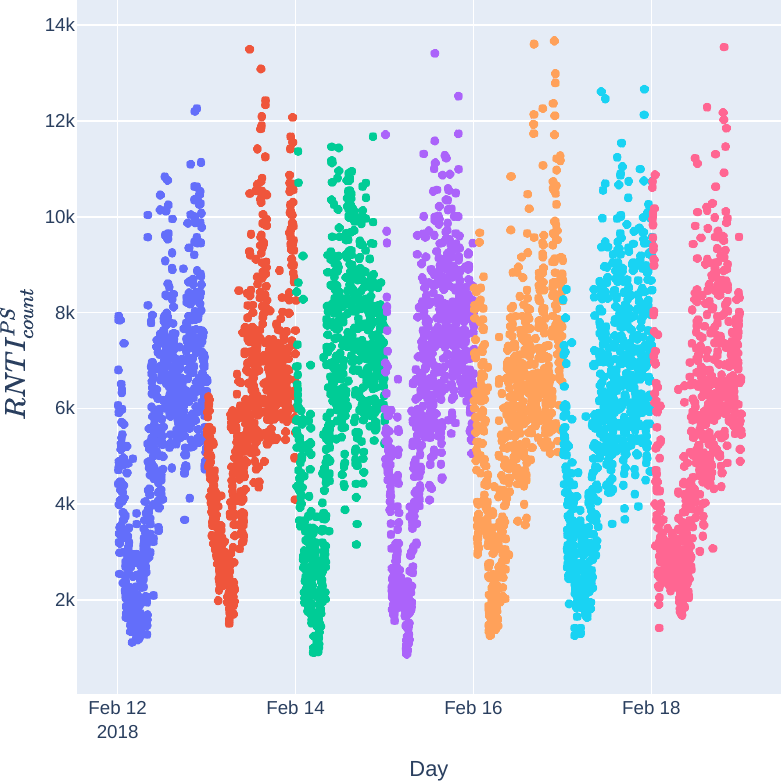}
		}
	\subfloat[\protect\label{subfig:PS_variables_rb_down}]{
		\includegraphics[width=.33\textwidth]{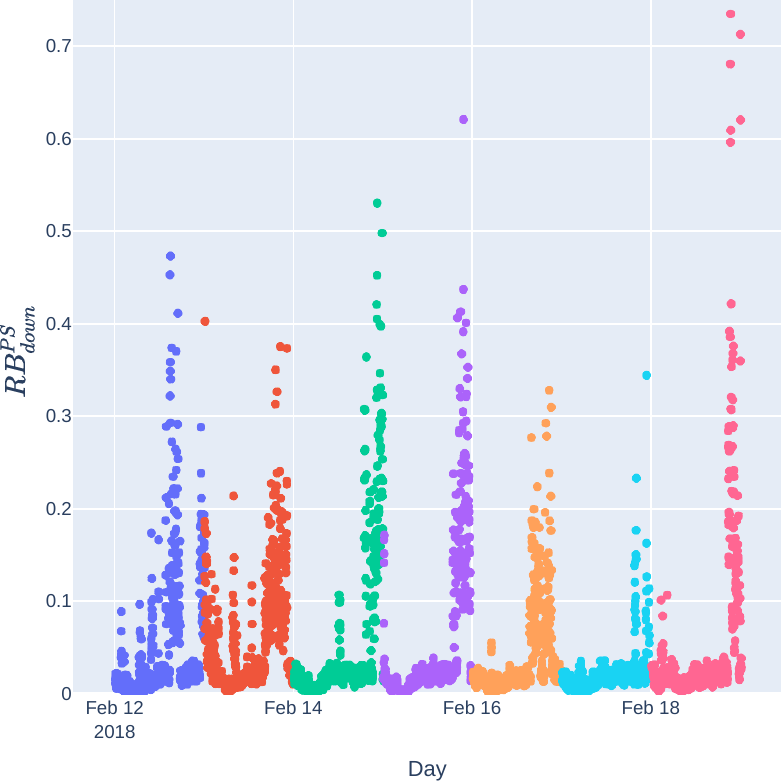}
		}
	\subfloat[\protect\label{subfig:PS_variables_rb_up}]{
		\includegraphics[width=.33\textwidth]{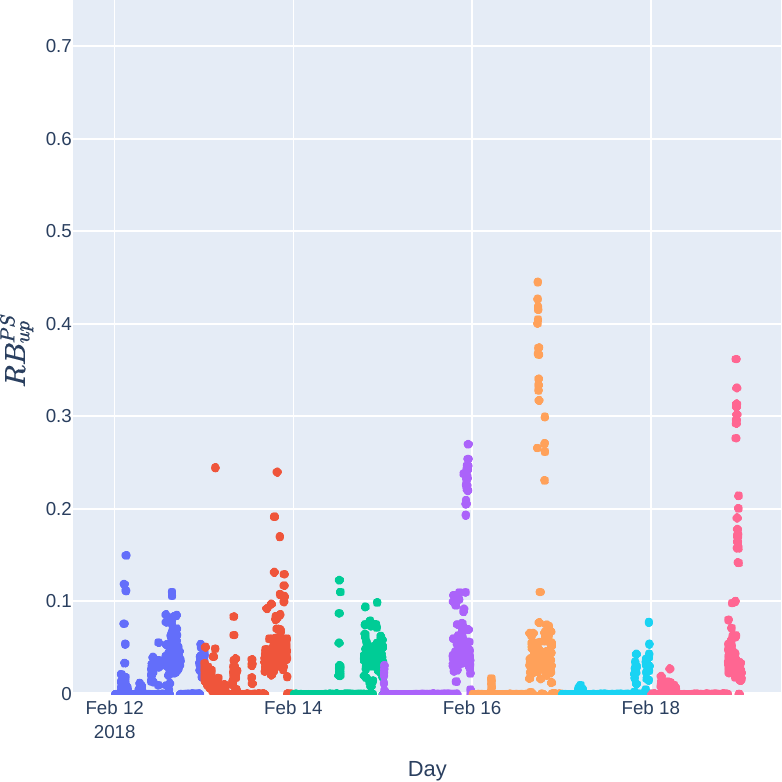}
		}
	\caption{Variables time series on a sample of 7 days of the PS dataset for \textit{RNTI}$_{count}$ (a), \textit{RB}$_{down}$ (b) and \textit{RB}$_{up}$ (c). Each color corresponds to a day of the week, starting from Monday which is the first represented with blue color.}
	\label{fig:PS_variables}
\end{figure*}

Fig.~\ref{fig:PS_variables} displays a \mbox{one-week} snapshot of the dataset variables. In particular, the evolution in time for PS data is outlined for \textit{RNTI}$_{count}$, \textit{RB}$_{down}$ and \textit{RB}$_{up}$, respectively. The data is characterized by a clear daily periodicity. This daily pattern is also evident across all other datasets, with slight differences appearing only in EB, where higher peaks occur on weekend nights, reflecting the popularity of the El Born district as a nightlife center in the city.

\begin{figure*}[h!]
	\centering
\	\includegraphics[width=0.99\textwidth]{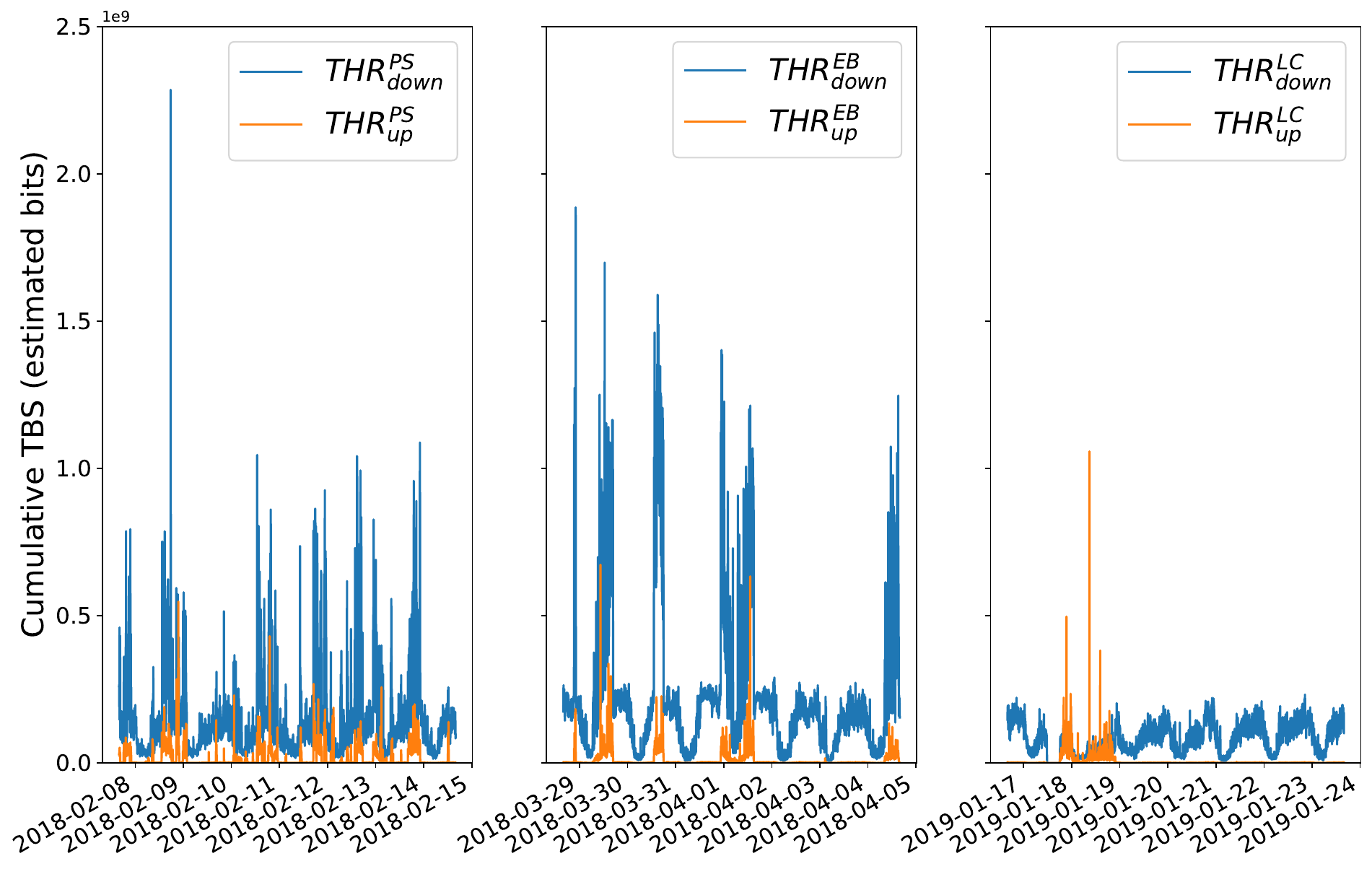}
	\caption{\textit{THR}$_{down}$ and \textit{THR}$_{up}$ time series by date for the three datasets on a sample of 7 days. The shared y-axis unit of measurement is billions of bits.}
	\label{fig:trafficComparison}
\end{figure*}

Finally, the impact on the scale proportion and the dynamics between the downlink and uplink channels of the three selected datasets is analyzed. Here, human behavior in the neighborhoods where the data was collected is of key importance: Poble-sec is mainly a residential area, while El Born boasts a lively nightlife and Les Corts hosts the Camp Nou football stadium. For this analysis, the total traffic demand as a function of the actual data transmitted is evaluated. Driven by this aim, we define \textit{THR}$_{down}$ and \textit{THR}$_{up}$ as \mbox{a-posteriori} estimation of the cumulative Transport Block Size (TBS) in bits in downlink and uplink, respectively, evaluated according to 3GPP specifications \cite{TS36213}. These metrics are reported in Fig.~\ref{fig:trafficComparison} for the different datasets during a week, for the sake of readability. The patterns between the uplink and downlink channels are similar for the PS and EB datasets; however, the range of values significantly differs, being \textit{THR}$_{up}$ lower than \textit{THR}$_{down}$. Alternatively, for LC, \textit{THR}$_{down}$ is much smaller with respect to the other databases, and \textit{THR}$_{up}$ shows peaks higher than \textit{THR}$_{down}$. These peaks correspond to periods when football matches occur at the Camp Nou stadium, located near the base station. We refer to our previous papers \cite{AnomaliesTrinh} and \cite{Pelati2022} for a more comprehensive study on mobile traffic analysis during these events.

\section{Traffic Prediction Models}
\label{sec:models}

\subsection{Problem Statement}\label{subsec:problem_statement}

Let $\mathcal{T}$ be the total measurement period of a given dataset. For every time $t \in \mathcal{T}$, $\bm{x}(t)$ is defined as the vector containing the $m$ input features. Note that $\mathcal{T}$ changes for each dataset, and this formulation is intended to be general for all of them. Similarly, the time interval $[t, t+1]$ is 2 minutes for every dataset, as justified in Section~\ref{sec:datasets}. 
The prediction model is organized as follows. Let $p$ be the number of input observations (samples), in terms of the number of dataset entries, constituting a single input instance, i.e., each input frame of the models will have shape $p \times m$: $\{\bm{x}(t), \bm{x}(t+1),..., \bm{x}(t+p-1)\} $.
Let $\Delta n$ be the time of the output frame to be predicted in terms of the number of samples with respect to the input frame, i.e., for a specific input frame $\bm{x}(t)$ the prediction will be the vector of $q$ variables $\bm{y}(t+\Delta n)$. 
The scenarios with $p \in \{10, 15, 20\}$ and $\Delta n \in \{0, 4, 9, 14\}$ have been studied; in other words, respectively, $20$, $30$ and $40$ minutes of past data are exploited to predict samples expected to occur $2$, $10$, $20$ and $30$ minutes after.

Given the multiple variables of $\bm{x}(t)$, \textit{data normalization} is performed for all the columns in each dataset in the interval $[-1,1]$ to keep the data symmetric with respect to the activation functions of the DL models used (i.e., the tanh). Moreover, \textit{data windowing} is applied to the sequence $\bm{x}(t)$, which is split and grouped using a \mbox{fixed-length} window of $p$ samples. The window is moved each time by one step. The value of $p$ defines how many time lags are processed by the DL models.
The \mbox{multi-variate} sequence can be expressed as $\left[ \bm{x}(1),\bm{x}(2), \dots, \bm{x}(T) \right]$, where $T$ is the cardinality of $\mathcal{T}$, i.e., $T = |\mathcal{T}|$. After the split, $N = T-p+1$ sequences $\mathbf{x}(n)$, $n \in [1,N]$ are available:
$$\mathbf{x}(1) = \left[ \bm{x}(1),\bm{x}(2), \dots, \bm{x}(p) \right]$$
$$\mathbf{x}(2) = \left[ \bm{x}(2), \dots, \bm{x}(p+1) \right]$$
$$\mathbf{x}(N) = \left[ \bm{x}(N), \dots, \bm{x}(T) \right]$$
A sequence $\mathbf{x}(n)$ has length $p$ and each of its elements is \mbox{$m$-dimensional}. Hereafter, we refer to the sequence $\mathbf{x}$ as \textit{samples}. Then, we define $\mathbf{X}$ as the \mbox{three-dimensional} matrix which contains $N$ sequences  of $\mathbf{x}$. The matrix $\mathbf{X}$ has dimension $N \times p \times m$ and serves as the input \textit{tensor} to the DL algorithms.

The multivariate input and output of the prediction model have been designed based on the EDA in Section~\ref{sec:datasets} and driven by the physical meaning of the features.
The resulting input features are the following $m=5$ variables: \textit{RNTI}$_{count}$, \textit{RB}$_{down}$, \textit{RB}$_{up}$, \textit{MCS}$_{down}$ and \textit{MCS}$_{up}$. 
Regarding the output, $q=5$ variables are exploited in the prediction models because of their valuable meaning for network operators, namely \textit{RNTI}$_{count}$, \textit{RB}$_{down}$, \textit{RB}$_{up}$, \textit{THR}$_{down}$ and \textit{THR}$_{up}$. The reason behind this choice is that these variables represent key performance indicators for network management, as detailed in Section~\ref{sec:background}.

In what follows, different DL architectures for the mobile traffic prediction problem are tailored. In particular, the proposed models are based on RNN and CNN architectures to exploit the temporal characteristics of the datasets under study. 
The proposal accounts for both \mbox{stand-alone} models and the DTL paradigm.

\subsection{RNN Model}\label{subsec:rnn_model}

Among the several RNN DL models, Gated Recurrent Units (GRUs) are adopted due to their high effectiveness and greater simplicity compared to other popular architectures, such as the LSTM cells \cite{Cahuantzi_2023}. 

\begin{figure}[h!]
    \centering
    \includegraphics[width=\linewidth]{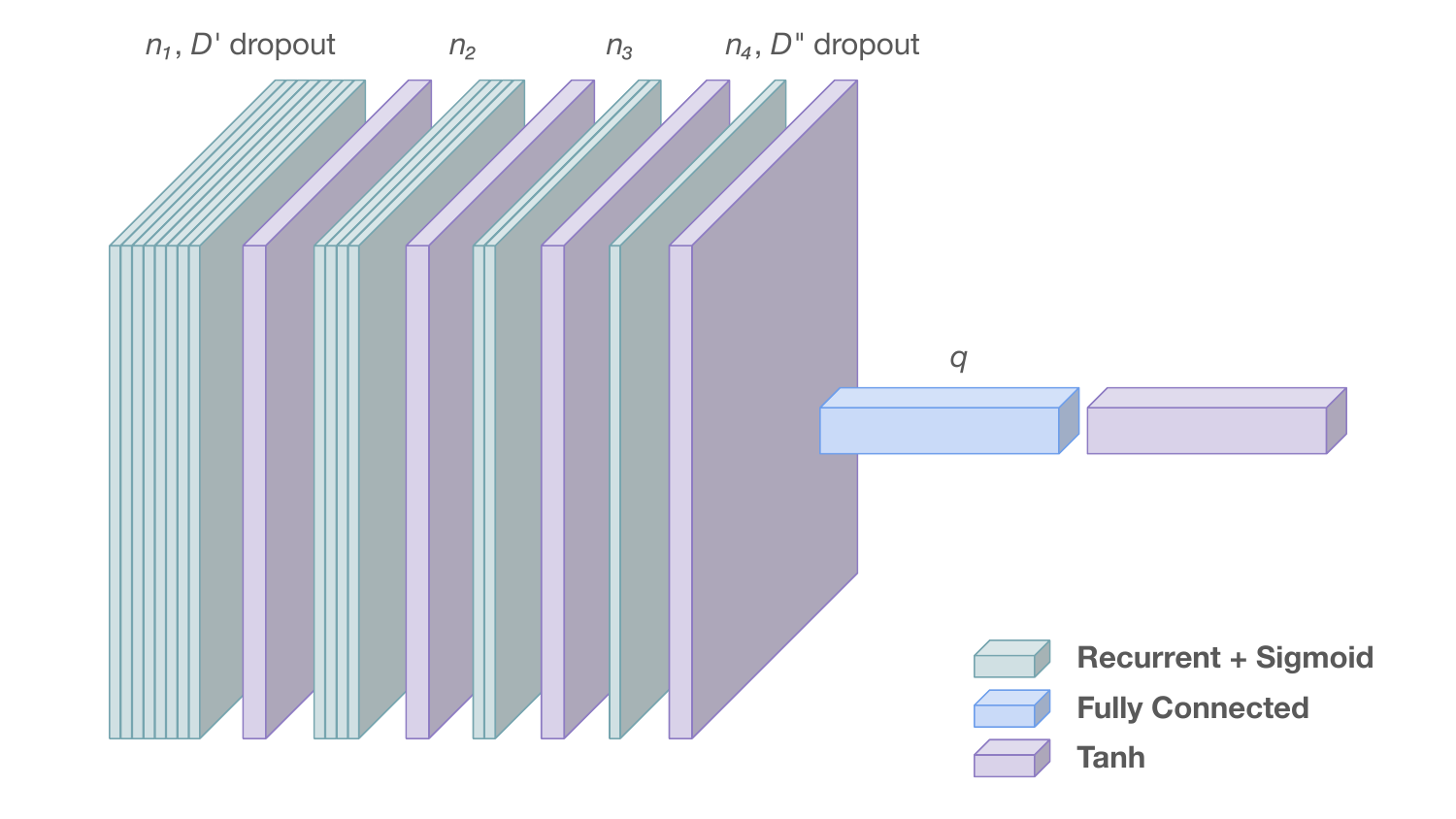}
    \caption{RNN model's structure graphical representation.}
    \label{fig:RNN-structure-graphic}
\end{figure}

\begin{figure}[h!]
    \centering
    \includegraphics[width=\linewidth]{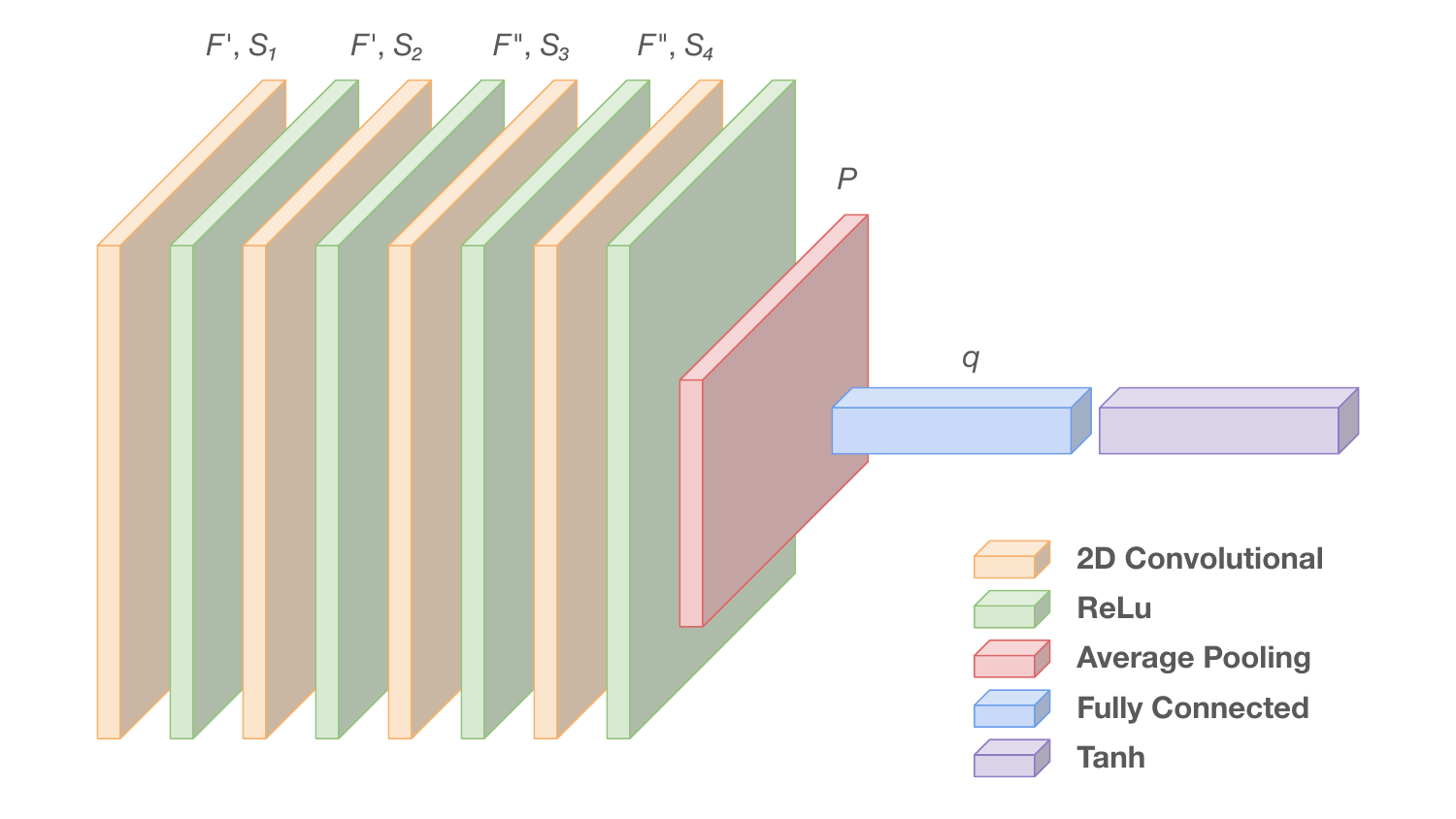}
    \caption{CNN model's structure graphical representation.}
    \label{fig:CNN-structure-graphic}
\end{figure}

After a comprehensive assessment of different configurations for the network structure, the architecture shown in Fig.~\ref{fig:RNN-structure-graphic} proved to be the best for the considered task.
The intuition behind its effectiveness is based on the aim to create \mbox{higher-level} abstractions and capture more nonlinearity between the data through the stacked layers. Moreover, the intent is to enforce the data correlation with the previous inputs, which is provided by the structure of each GRU cell.
In detail, after the input layer, a first recurrent layer is inserted, containing $n_1$ GRU cells with sigmoid activation for the input, forget and output gates, and hyperbolic tangent activation for the hidden and output states; dropout regularization with probability parameter $D'$ is also implemented to prevent overfitting.
The second and third layers are based on GRU with the same activation function of the first, and contain respectively $n_2$ and $n_3$, whereas the fourth layer contains $n_4$ such GRU cells and dropout regularization with probability parameter $D''$.
At the end, the tensor is flattened by introducing the final dense layer containing $q$ neurons, and a hyperbolic tangent activation function is added to obtain the prediction output. Note that the final activation is selected to meet the rescaling interval of the normalized output data ($[-1,1]$), whereas the choice of adding dropout regularization to the first and last layers of the network is motivated by \cite{dropoutPositionRNN}.

The best performing RNN hyperparameters are reported in Table~\ref{tab:hyperparameters}a, as a result of a \mbox{grid-search}.

\subsection{CNN Model}\label{subsec:cnn_model}

The second DL architecture is based on CNNs. As with RNNs, different network configurations have been tested, finally leading to the choice of the structure represented in Figure~\ref{fig:CNN-structure-graphic}.
To understand the strength of this architecture, each input feature can be thought of as an image, where each value corresponds to a pixel.
As a processed input goes deeper in the network, \mbox{higher-level} abstractions of it are extracted and nonlinearly correlated through the activation functions. In this case, the advantages consist of a significantly lower number of parameters to be trained and in a natural mitigation of the vanishing gradient problem affecting the RNN architectures. 
In detail, the input layer is followed by four \mbox{two-dimensional} convolutional layers, each accompanied by a Rectified Linear Unit (ReLU) activation layer, with a \mbox{non-decreasing} number of filters and different ad hoc kernel sizes. 
Leaky ReLU activation functions are chosen to solve the well-known problem of trained weights that permanently prevent a set of neurons from being activated, caused by ReLU functions' zero slope for negative \mbox{$x$-axis} values. 

The first two convolutional layers contain $F'$ filters, respectively of dimension $S_1$ and $S_2,$ while the third and fourth convolutional layers contain $F''$ filters of dimension is $S_3$ and $S_4.$
The filters' stride is equal to one in all the cases (one-step on the horizontal movements, and one-step row change on the vertical movements) since the dimensionality of the problem is quite small and no important information shall be lost.
After that, a two-dimensional average pooling layer provides a valuable summary of the extracted information along the temporal axis, thanks to the pooling size $P=[2,1]$. Finally, like in the RNNs, a flattening layer is inserted, followed by a final dense layer containing $q$ neurons with hyperbolic tangent activation to obtain the prediction. 
Note that, in the CNNs' case, the order of the features is rearranged as follows: \textit{RB}$_{down},$ \textit{RB}$_{up},$ \textit{RNTI}$_{count},$ \textit{MCS}$_{down},$ and \textit{MCS}$_{up}.$
The reason is to allow the network to exploit the \mbox{a-priori} uncorrelation of \textit{RNTI}$_{count}$ with all the other variables, given the local approach of kernels among the features in CNN models. 

The best performing CNN hyperparameters resulting from a \mbox{grid-search} are reported in Table~\ref{tab:hyperparameters}b.
The resulting best value of $S_1=[16,3]$ implies that a correlation of a higher number of variables from the first hidden layer can help the model to perform better. In particular, given the filters' size and the unitary stride, the \textit{RNTI}$_{count}$ variable is the only one to get directly involved with all the other features in the CNN convolutions since the layer closest to the input of the network. 
$S_2=[3,5]$, being the best choice for the second layer filters dimension, indicates that correlating all the five input variables three time steps at a time is the best \mbox{trade-off} to maximize the extracted information. Furthermore, $F''=32$, as the ideal number of filters for both the third and the fourth layers, suggests that too many filters are not needed when their size is chosen properly, with $S_3=[8,3]$ and $S_4=[4,3]$ in the described case.
Finally, $P=[2,1]$ confirms that the average pooling layer is useful to provide a summary of the convolutional layers' output along the temporal axis. However, it also reveals that it only needs to involve two rows to retrieve the relevant insights extracted from the internal layers and provide a reliable prediction.

\begin{table*}
	\centering
	\subfloat[RNN]{
		\begin{tabular}{cccccc}
			\multicolumn{6}{c}{} \\
			\hline
			$n_1$ & $n_2$ & $n_3$ & $n_4$ & $D'$ & $D''$ \\%
			$128$ & $64$ & $32$ & $16$ & $0$ & $0.2$ \\
			\hline
	\end{tabular}}
	\hspace{16mm}
	\subfloat[CNN]{
		\begin{tabular}{ccccccc}
			\multicolumn{7}{c}{} \\
			\hline
			$F'$ & $F''$ & $S_1$ & $S_2$ & $S_3$ & $S_4$ & $P$ \\%
			$16$ & $32$ & $[16,3]$ & $[3,5]$ & $[8,3]$ & $[4,3]$ & $[2,1]$ \\
			\hline
	\end{tabular}}
	\caption{DL models' optimized hyperparameters.}\label{tab:hyperparameters}
\end{table*}

\subsection{Deep Transfer Learning}\label{subsec:models_transfer}

The adopted DTL method is based on network-based deep transfer learning, which reuses part of the network that has been pre-trained in the source domain as part of the deep neural network used in the target domain~\cite{tan2018_DTLsurvey}. In particular, the \emph{frozen layers} approach is adopted, which consists of inhibiting the update of the weights of the neurons placed in different combinations of layers, and thus training the remaining part of the networks on the new data~\cite{iman2023review}. This approach is grounded on the assumption that, for both RNNs and CNNs, each layer extracts meaningful features from the output of the previous one. This implies that groups of neurons that are close (considering the adopted metric on the network graph) should be specialized in extracting similar types of information.

The idea is to perform new training only on the part of the network that is responsible for extracting the features devoted to describing the differences among the datasets. Thus, the common general information already obtained by the source model (and maintained by the frozen neurons) is preserved, while the model is adapting the prediction to the new input distribution. For example, in CNN architectures, the CNN layers closer to the input are devoted to extracting features from the given dataset and can be frozen in DTL. Alternatively, the fully connected layers before the output are responsible for classification and, thus, need to be trained with the target data.

A similar DTL solution is represented by \emph{fine-tuning}, where a pre-trained model is trained with the target data. Despite its effectiveness in DTL methods for many tasks and datasets in various fields, it still represents a computationally expensive solution, since it needs to train the whole ML network with the student's data.
Instead, the \emph{frozen layers} approach limits such an issue by training only a subset of the ML network.
\section{Experiments and results}
\label{sec:results}

The following section details the characteristics of the experiments performed and the numerical results of the different models. 
For the sake of clarity, the section is divided into five subsections.
In Section~\ref{subsec:results_detail} the model training setup is detailed. 
Section~\ref{subsec:results_FS} reports the performance of the \mbox{stand-alone} models on the PS, EB, and LC datasets, respectively. The usage of DTL and its performance is detailed in Section~\ref{subsec:results_transfer}.
Model complexity and the energy consumption of the studied models are discussed in Section~\ref{subsec:results_complexity}.
Finally, Section~\ref{subsec:results_SVR} contains the comparison with the benchmark SVR models.

\subsection{Model Training Setup}\label{subsec:results_detail}

Each dataset has been split into a training $\mathcal D_{t,i}$ and a validation set $\mathcal D_{v,i}$, each of them, in turn, divided into two additional different proportions ($i \in [1,2]$) to study the best threshold between the size of the training set $|\mathcal D_t|$ and the performance of each model. 
In particular, $\mathcal D_{t,1}^{\text{PS}}$ and $\mathcal D_{t,2}^{\text{PS}}$ contain respectively fourteen and \mbox{twenty-one} days of data in PS.
$\mathcal D_{t,1}^{\text{EB}}$ and $\mathcal D_{t,1}^{\text{LC}}$ contain, instead, five days of data, whereas $\mathcal D_{t,2}^{\text{EB}}$ and $\mathcal D_{t,2}^{\text{LC}}$ cover a \mbox{six-day} time period.
No significant difference was found between the two setups, besides a more consistent convergence of the loss function when training the models on the biggest training set. The results will hence be reported for $\mathcal D_{t,2}$ and $\mathcal D_{v,2}$ only, referred to as $\mathcal D_t$ and $\mathcal D_v$ in the rest of this work.

The models were trained and validated using an Ubuntu 18.04 High-Performance Cluster, with 2 Intel(R) Xeon(R) Gold 6230 CPU @ 2.10GHz, 187 GB of RAM, and four NVIDIA GeForce RTX 2080 Ti GPUs. The DL algorithms were implemented in Python, using the \texttt{Keras} library on top of the \texttt{Tensorflow} backend. For the benchmark SVR, the popular \texttt{Scikit-Learn} library was used.

The regression performance was assessed by looking at the Mean Squared Error (MSE) between the true validation data and the predictions, divided by the cardinality of the validation set $|\mathcal{D}_v|.$
This allows to fairly compare models, which have been trained on different portions of the same dataset.  
To achieve reliable results, the experiments on each model were run three times, and the resulting output metrics were averaged.

For the number of training epochs, a unique \mbox{upper-bound} was adopted for all cases, regardless of the trained structure and the dataset. This choice aims to reduce the manual intervention in the training process, avoiding setting a customized hyperparameter for each case.
The number of epochs was set to 30, also in the DTL framework, where the needed epochs are even fewer. Average values for the actual necessary number of training epochs will be reported in Section~\ref{subsec:results_complexity}.  Source code for the proposed strategies, along with the experiments' configuration and the code to reproduce the experiments is available online\footnote{\url{https://gitlab.cttc.es/supercom/traffic-prediction}}. 

\subsection{Stand-alone models}\label{subsec:results_FS}
As a first step of the analysis, RNN and CNN models are trained on the data collected from the three locations in Barcelona (PS, EB, and LC). 
The results in terms of MSE for the datasets $\mathcal D_t^{\text{PS}}$, $\mathcal D_t^{\text{EB}}$, and $\mathcal D_t^{\text{LC}}$ are reported in Table~\ref{tab:MSE_PS21}, Table~\ref{tab:MSE_EB6} and Table~\ref{tab:MSE_LC6}, respectively. 
The results are detailed by varying the input window of observation $p$ in the columns and the time interval from the input and the output tensor $\Delta n$ in the rows. 
The lowest MSE between the CNN and RNN models performing the same task has been highlighted in yellow to improve the readability of the tables. 

\begin{table*}
	\centering
	$\begin{array}{c|cc|cc|cc|c}
	\Delta n \backslash p & \multicolumn{2}{c|}{10} & \multicolumn{2}{c|}{15} & \multicolumn{2}{c|}{20} & \\
	\hline
	0 & 0.0091                         & \cellcolor[HTML]{FFF4BD}0.0088 & \cellcolor[HTML]{FFF4BD}0.0087 & 0.0089                         & \cellcolor[HTML]{FFF4BD}0.0088 & 0.0090 & \text{MSE} \\
	\hline
	4 & \cellcolor[HTML]{FFF4BD}0.0153 & 0.0156                         & 0.0149                         & \cellcolor[HTML]{FFF4BD}0.0148 & \cellcolor[HTML]{FFF4BD}0.0151 & 0.0155 & \text{MSE} \\
	\hline
	9 & \cellcolor[HTML]{FFF4BD}0.0175 & 0.0176                         & 0.0183                         & \cellcolor[HTML]{FFF4BD}0.0174 & 0.0184                         & \cellcolor[HTML]{FFF4BD}0.0180 & \text{MSE} \\
	\hline
	14 & 0.0197                         & \cellcolor[HTML]{FFF4BD}0.0196 & \cellcolor[HTML]{FFF4BD}0.0193 & \cellcolor[HTML]{FFF4BD}0.0193 & 0.0198                         & \cellcolor[HTML]{FFF4BD}0.0192 & \text{MSE} \\
	\hline
	& \text{RNN} & \text{CNN} & \text{RNN} & \text{CNN} & \text{RNN} & \text{CNN} & \\
	\end{array}$
	\caption{MSE of the stand-alone models trained on $\mathcal D_t^{\text{PS}}$.}\label{tab:MSE_PS21}
\end{table*}

\begin{table*}
	\centering
	$\begin{array}{c|cc|cc|cc|c}
	\Delta n \backslash p & \multicolumn{2}{c|}{10} & \multicolumn{2}{c|}{15} & \multicolumn{2}{c|}{20} & \\
	\hline
	0 & \cellcolor[HTML]{FFF4BD}0.0062 & 0.0065                         & \cellcolor[HTML]{FFF4BD}0.0062 & 0.0070                         & \cellcolor[HTML]{FFF4BD}0.0063 & 0.0068 & \text{MSE} \\
	\hline
    4 & 0.0083                         & \cellcolor[HTML]{FFF4BD}0.0079 & \cellcolor[HTML]{FFF4BD}0.0081 & \cellcolor[HTML]{FFF4BD}0.0081 & 0.0084                         & \cellcolor[HTML]{FFF4BD}0.0076 & \text{MSE} \\
	\hline
    9 & 0.0081                         & \cellcolor[HTML]{FFF4BD}0.0079 & \cellcolor[HTML]{FFF4BD}0.0081 & 0.0088                         & \cellcolor[HTML]{FFF4BD}0.0078 & 0.0087 & \text{MSE} \\
	\hline
    14 & 0.0090                         & \cellcolor[HTML]{FFF4BD}0.0089 & 0.0099                         & \cellcolor[HTML]{FFF4BD}0.0092 & \cellcolor[HTML]{FFF4BD}0.0089 & 0.0090 & \text{MSE} \\
	\hline
	& \text{RNN} & \text{CNN} & \text{RNN} & \text{CNN} & \text{RNN} & \text{CNN} & \\
	\end{array}$
	\caption{MSE of the stand-alone models trained on $\mathcal D_t^{\text{EB}}.$}\label{tab:MSE_EB6}%
\end{table*}

\begin{table*}
	\centering
	$\begin{array}{c|cc|cc|cc|c}
	\Delta n \backslash p & \multicolumn{2}{c|}{10} & \multicolumn{2}{c|}{15} & \multicolumn{2}{c|}{20} & \\
	\hline
    0 & \cellcolor[HTML]{FFF4BD}0.0089 & \cellcolor[HTML]{FFF4BD}0.0089 & \cellcolor[HTML]{FFF4BD}0.0086 & 0.0090                         & \cellcolor[HTML]{FFF4BD}0.0089 & \cellcolor[HTML]{FFF4BD}0.0089 & \text{MSE} \\
	\hline
    4 & 0.0144                         & \cellcolor[HTML]{FFF4BD}0.0141 & \cellcolor[HTML]{FFF4BD}0.0136 & 0.0141                         & \cellcolor[HTML]{FFF4BD}0.0137 & 0.0141 & \text{MSE} \\
	\hline
    9 & \cellcolor[HTML]{FFF4BD}0.0164 & \cellcolor[HTML]{FFF4BD}0.0164 & 0.0166                         & \cellcolor[HTML]{FFF4BD}0.0159 & \cellcolor[HTML]{FFF4BD}0.0161 & 0.0165 & \text{MSE} \\
	\hline
    14 & 0.0187                         & \cellcolor[HTML]{FFF4BD}0.0184 & 0.0187                         & \cellcolor[HTML]{FFF4BD}0.0183 & \cellcolor[HTML]{FFF4BD}0.0190 & \cellcolor[HTML]{FFF4BD}0.0190 & \text{MSE} \\
	\hline
	& \text{RNN} & \text{CNN} & \text{RNN} & \text{CNN} & \text{RNN} & \text{CNN} & \\
	\end{array}$
	\caption{MSE of the stand-alone models trained on $\mathcal D_t^{\text{LC}}.$}\label{tab:MSE_LC6}%
\end{table*}

In most cases, the MSE increases with $\Delta n$, as expected. For example, the MSE for $\Delta n = 14$ doubles that of $\Delta n = 0$ with datasets $\mathcal D_t^{\text{PS}}$ and $\mathcal D_t^{\text{LC}}$. Interestingly, RNNs in general perform better with small values of $\Delta n,$ whereas CNNs outperform for higher values. This phenomenon can be motivated by CNNs being able to better represent the temporal structure of input signals for longer time horizons, which leads to better predictions for higher $\Delta n$.

Another noticeable trend is that MSE values are stable with $p$ for both RNNs and CNNs. 
This result suggests that information temporally distant from the models' output does not contribute to improving prediction accuracy, which can therefore be achieved using only the past 20 minutes ($p=20$).
This property will be confirmed by the XAI tools in Section~\ref{sec:explainability}.

\subsection{DTL models}\label{subsec:results_transfer}

For this analysis, the larger PS dataset is considered as the teacher, who transfers the knowledge obtained in the \mbox{stand-alone} model to the other two (student) datasets (EB and LC). 
The approach in \cite{tan2018_DTLsurvey} is followed, which consists of building the new (student) model by selecting a subset of layers of the teacher model (those closer to the input) as trained on the teacher dataset (an operation called \emph{layer freezing}). The remaining layers, closer to the output, are re-trained based on the new (student) data.
The rationale is based on the assumption that each layer is able to extract meaningful and increasingly specific features.
Therefore, layers closer to the output are supposed to be those extracting features for that specific learning task, namely the traffic prediction of the student in the new location. 
Consequently, in this work, the model behavior is studied when \mbox{re-training} part of the layers close to the output. In particular, all the possible combinations in \mbox{re-training} each of the last three layers of the two DL models are considered. 
One of these combinations also includes the teacher model trained with the PS dataset (i.e., the student model with all frozen layers from the teacher). This instance is an extreme example of knowledge transfer without any extra training cost.

Table~\ref{tab:MSE_EB6_T} shows the lowest MSE of the models among the different freezing combinations using DTL on EB data, using $\mathcal D_t^{\text{PS}}$ to train the teacher model. Table~\ref{tab:MSE_LC6_T} provides the same results for the LC dataset. The blue color highlights which of the two considered DL models returns the lowest MSE.

\begin{table*}
	\centering
	$\begin{array}{c|cc|cc|cc|c}
	\Delta n \backslash p & \multicolumn{2}{c|}{10} & \multicolumn{2}{c|}{15} & \multicolumn{2}{c|}{20} & \\
	\hline
	0  & \cellcolor[HTML]{CEE5ED}0.0060 & \cellcolor[HTML]{CEE5ED}0.0060 & \cellcolor[HTML]{CEE5ED}0.0059 & 0.0060                         & 0.0060                         & \cellcolor[HTML]{CEE5ED}0.0059 & \text{MSE} \\
	\hline
	4 & \cellcolor[HTML]{CEE5ED}0.0077                         & 0.0079 & 0.0078                         & \cellcolor[HTML]{CEE5ED}0.0074 & \cellcolor[HTML]{CEE5ED}0.0077                      & 0.0078 & \text{MSE} \\
	\hline
	9 & 0.0081                         & \cellcolor[HTML]{CEE5ED}0.0078 & 0.0085                         & \cellcolor[HTML]{CEE5ED}0.0079 & \cellcolor[HTML]{CEE5ED}0.0080 & 0.0082 & \text{MSE} \\
	\hline
	14 & 0.0092                         & \cellcolor[HTML]{CEE5ED}0.0089 & 0.0096                         & \cellcolor[HTML]{CEE5ED}0.0088 & 0.0096                         & \cellcolor[HTML]{CEE5ED}0.0091 & \text{MSE} \\
	\hline
	& \text{RNN (T)} & \text{CNN (T)} & \text{RNN (T)} & \text{CNN (T)} & \text{RNN (T)} & \text{CNN (T)} & \\
	\end{array}$
	\caption{MSE of the best transferred models for each of the considered cases, trained on $\mathcal D_t^{\text{EB}}.$}\label{tab:MSE_EB6_T}%
\end{table*}

\begin{table*}
	\centering
	$\begin{array}{c|cc|cc|cc|c}
	\Delta n \backslash p & \multicolumn{2}{c|}{10} & \multicolumn{2}{c|}{15} & \multicolumn{2}{c|}{20} & \\
	\hline
	0  & \cellcolor[HTML]{CEE5ED}0.0084 & \cellcolor[HTML]{CEE5ED}0.0084 & \cellcolor[HTML]{CEE5ED}0.0085 & \cellcolor[HTML]{CEE5ED}0.0085 & \cellcolor[HTML]{CEE5ED}0.0085 & \cellcolor[HTML]{CEE5ED}0.0085 & \text{MSE} \\
	\hline
	4 & \cellcolor[HTML]{CEE5ED}0.0136 & \cellcolor[HTML]{CEE5ED}0.0136 & 0.0135 & \cellcolor[HTML]{CEE5ED}0.0134 & \cellcolor[HTML]{CEE5ED}0.0135 & \cellcolor[HTML]{CEE5ED}0.0135 & \text{MSE} \\
	\hline
	9 & \cellcolor[HTML]{CEE5ED}0.0162                         & 0.0164 & 0.0163                         & \cellcolor[HTML]{CEE5ED}0.0160 & 0.0161 & \cellcolor[HTML]{CEE5ED}0.0160 & \text{MSE} \\
	\hline
	14 & 0.0183                         & \cellcolor[HTML]{CEE5ED}0.0180 & 0.0184                         & \cellcolor[HTML]{CEE5ED}0.0181 & 0.0181 & \cellcolor[HTML]{CEE5ED}0.0180 & \text{MSE} \\
	\hline
	& \text{RNN (T)} & \text{CNN (T)} & \text{RNN (T)} & \text{CNN (T)} & \text{RNN (T)} & \text{CNN (T)} & \\
	\end{array}$
	\caption{MSE of the best transferred models for each of the considered cases, trained on $\mathcal D_t^{\text{LC}}.$}\label{tab:MSE_LC6_T}%
\end{table*}

The trends observed for the \mbox{stand-alone} models with respect to $\Delta n$ and $p$ are maintained. However, CNN models show higher accuracy in twice as many cases as RNNs. Specifically, CNN and RNN models perform equally well in 6 cases, while CNN models outperform in 13 cases and RNN models in 5.
In fact, GRU models extract task-specific features already from the first layers.
Therefore, GRU re-trained layers are not able to adapt to the change of domain as accurately as CNNs. This behavior could depend on \mbox{CNN-based} models holding a more hierarchical knowledge distillation architecture, in which the deepest layers are specialized for the specific prediction task to be performed. 

Interestingly, no significant difference can be identified when applying DTL from a teacher model using 
$\mathcal D_{t,1}^{\text{PS}}$ or $\mathcal D_{t,2}^{\text{PS}}$. 
This result suggests that fourteen days of data are enough to train the ML model for extracting the traffic patterns and being able to transfer them to a student model.

Finally, DTL enables an accuracy improvement in more than $85\%$ of the studied cases ($41$ out of $48$) with respect to the \mbox{stand-alone} counterpart. 
This result confirms that knowledge transfer is a valid method to accurately train models when available data is scarce (as in the case of EB and LC).

\subsection{Complexity analysis and Energy Cost}\label{subsec:results_complexity}

The computational complexity of the proposed solutions depends on four factors: the dataset size, the model structure, the number of training epochs, and the number of trainable parameters in the model. As a proxy for the complexity, an analysis based on those factors has been conducted, following the Green AI principles~\cite{Schwartz2019}.
In particular, the average training time of a single model among the different values of $\Delta n$ is provided to offer a more reliable estimation of the required resources, as the temporal distance of the prediction from the input instances does not affect the training time.
Moreover, the drained total training energy is calculated using the online tool Green Algorithms \cite{greenAlgorithms}.
Results are reported in Table~\ref{table:energy-results3} for both \mbox{stand-alone} and DTL cases and using $\mathcal D_t^{\text{EB}}.$

RNN models contain a total of $100\,981$ parameters. Instead, \mbox{CNN-based} models have a number of parameters varying with $p,$ namely $33\,285$, $34\,885$ or $37\,285$, with $p$ equal to $10,$ $15$ or $20$, respectively. In fact, for CNNs with fixed filter sizes in each layer, the number of input values to the final dense layer depends on the number of the filters' strides. This is reflected in a linear increase in the number of weights for each output neuron. Complexity metrics vary very little for CNNs as a function of $p$, so the table contains only values for $p=10$.
The number of parameters for \mbox{CNN-based} \mbox{stand-alone} models is roughly \mbox{one-third} that of the corresponding RNNs.
This highlights the intrinsic advantage of the convolutional layers, which is reflected in the reduced number of training epochs to reach the minimum loss and the training time that is 10 times lower than RNN.
In turn, this difference impacts the energy footprint, which is 10 times lower for CNNs.

\begin{table*}[tb]%
  \caption{Model training complexity and energy consumption for both stand-alone and DTL models, trained on $\mathcal D_t^{\text{EB}}$, $p=10$.}
\begin{center}
\begin{tabular}{c|c|c|c|c}
\toprule
& \multicolumn{2}{c | }{Stand-alone} & \multicolumn{2}{c}{DTL} \\
&  RNN & CNN & RNN & CNN \\
\midrule
\hline
\text{N. of parameters} & 100\,981 & 33\,285 & 9\,449.25 & 13\,321.25\\
\hline
\text{N. of training epochs} & 13 & 9 & 6 & 8 \\
\hline
\text{Training time [s]} & 74.409 & 3.908 & 3.804 & 1.586 \\
\hline
\text{Energy drained [Wh]} & 19.8 & 1.0 & 1.0 & 0.4 \\
\hline
\text{\% of saved energy by DTL} & - & - & 94.95 & 60.00 \\
\hline
\bottomrule
\end{tabular}
\end{center}
\label{table:energy-results3}
\end{table*}

Considering the DTL models, RNNs need a smaller number of epochs to reach the loss minimum. However, CNN models require less than half of the RNNs training time. This translates into a lower energy footprint of CNNs, which is 40\% with respect to RNNs, as with \mbox{stand-alone} models.
It can be argued that the energy footprint figures are tightly related to the training time, as experienced both for DTL and \mbox{stand-alone}; thus, CNNs represent the most \mbox{energy-efficient} model.
On the other hand, the gain of DTL in RNN with respect to \mbox{stand-alone} models is 95\%, whereas the CNN architectures reach 60\%, as detailed in Table~\ref{table:energy-results3}. Consequently, DTL has been able to reduce energy usage more for the ML architecture that presents the highest complexity (RNN).

Concluding, in addition to substantially decreasing the used energy, DTL also enables an accuracy improvement in $41$ out of $48$ (85\%) studied cases with respect to the \mbox{stand-alone} counterpart. 

\subsection{Comparison with Benchmark}\label{subsec:results_SVR}

In this subsection, the proposed RNN and CNN-based models are compared with a benchmark based on SVR~\cite{support1997_SVR}. 
A transformation based on the Radial Basis Function (RBF) kernel \cite{WANG2003643} is applied to handle the nonlinearity of the datasets. 
The \mbox{best-performing} hyperparameters are considered, which are the result of a \mbox{grid search} carried out considering several values of the margin width $\epsilon$ and the regularization parameter $C$.

\begin{table*}
	\centering
	$\begin{array}{c|c|c|c|c|c|c|c|c|c|c}
	& \multicolumn{3}{c|}{\text{Poble Sec}} & \multicolumn{3}{c|}{\text{El Born}} & \multicolumn{3}{c|}{\text{Les Corts}} & \\
	\hline
	\Delta n \backslash p & 10 & 15 & 20 & 10 & 15 & 20 & 10 & 15 & 20 & \\
	\hline
    0 & \cellcolor[HTML]{EFBE7D}0.0086 & 0.0087                         & \cellcolor[HTML]{EFBE7D}0.0087 & \cellcolor[HTML]{EFBE7D}0.0061 & 0.0062                         & 0.0063 & \cellcolor[HTML]{EFBE7D}0.0085 & 0.0086                         & \cellcolor[HTML]{EFBE7D}0.0086 & \text{MSE} \\
	\hline
    4 & 0.0153                         & \cellcolor[HTML]{C1E1C1}0.0152 & \cellcolor[HTML]{C1E1C1}0.0153 & \cellcolor[HTML]{FAA0A0}0.0075 & \cellcolor[HTML]{EFBE7D}0.0077 & 0.0076 & \cellcolor[HTML]{EFBE7D}0.0138 & \cellcolor[HTML]{C1E1C1}0.0137 & 0.0137 & \text{MSE} \\
	\hline
    9 & \cellcolor[HTML]{C1E1C1}0.0181 & \cellcolor[HTML]{C1E1C1}0.0181 & \cellcolor[HTML]{C1E1C1}0.0181 & \cellcolor[HTML]{C1E1C1}0.0080 & \cellcolor[HTML]{EFBE7D}0.0080 & \cellcolor[HTML]{C1E1C1}0.0081 & 0.0164                         & \cellcolor[HTML]{C1E1C1}0.0163 & \cellcolor[HTML]{C1E1C1}0.0164 & \text{MSE} \\
	\hline
    14 & \cellcolor[HTML]{C1E1C1}0.0202 & \cellcolor[HTML]{C1E1C1}0.0202 & \cellcolor[HTML]{C1E1C1}0.0204 & \cellcolor[HTML]{CFCFC4}0.0087 & \cellcolor[HTML]{CFCFC4}0.0088 & \cellcolor[HTML]{C1E1C1}0.0091 & 0.0184                         & \cellcolor[HTML]{C1E1C1}0.0186 & \cellcolor[HTML]{EFBE7D}0.0189 & \text{MSE} \\
	\end{array}$
	\caption{MSE of the benchmark SVR models for each of the considered cases, trained on $\mathcal D_t^{\text{PS}},$ $\mathcal D_t,^{\text{EB}}$ and $\mathcal D_t^{\text{LC}}.$}\label{tab:MSE_21-6-6_SVR}%
\end{table*}

Table~\ref{tab:MSE_21-6-6_SVR} contains the results of the SVR model in terms of MSE as a function of $\Delta n$ and $p$. 
For PS, which is used for \mbox{stand-alone} models only, the corresponding entries are highlighted in orange where the SVR model performs better, in green when performing worse, and left blank (no color) when the performance is the same. 
For the EB and LC, used for both \mbox{stand-alone} and DTL models, a different color mapping is adopted. A \mbox{red-filled} cell indicates better MSE of SVR compared to both the \mbox{stand-alone} and DTL models; a gray-filled cell means an MSE equal to the DTL model, but better than the \mbox{stand-alone}; a blank cell implies an MSE equal to the \mbox{stand-alone} model, but worse than the DTL; a \mbox{green-filled} cell indicates an MSE worse than both the \mbox{stand-alone} and DTL models. 

SVR performs better for low values of $p$ and $\Delta n$. This is due to the fact that in such cases the prediction task is easier, since the amount of input data is the smallest among all the studied cases and the prediction is closer in time to the input frame.
This can be spotted by looking at the high concentration of green-filled cells for $\Delta n$ equal to $9$ and $14$ and orange-filled cells for $\Delta n$ equal to $0$.
Similarly, SVR performance is often comparable to or better than DL models in EB and LC due to the smaller cardinality of these training sets. This is highlighted in the table by the high number of orange and red-filled cells in the columns of EB and LC.

Therefore, SVR achieves higher accuracy than DL in 9 cases whereas DL outperforms SVR in 14 cases out of a total of 39 (with both achieving the same accuracy in 16 cases).
However, DL models outperform SVR when i) the cardinality of the training set is larger, ii) the dimensionality of the input data increases ($p$) and, iii) the prediction is distant from the input ($\Delta n$). This represents a limit in the generalization properties of SVR applied to big data problems for mobile traffic prediction tasks, as the opposite of the proposed DL models, which scale better in such scenarios. 

Moreover, SVR presents significant limitations in the applicability of DTL techniques such as layer freezing. In the literature, approaches like Transfer Component Analysis (TCA)~\cite{transferSVR} can be applied, though they rely on jointly processing the teacher and student datasets. Thus, using SVR does not allow the separation of the training phase into two (teacher and student learning).
\section{Models explainability}
\label{sec:explainability}

Two fairly known XAI algorithms are tailored, namely SmoothGrad \cite{SmoothGrad} and LRP \cite{LRP}, to provide interpretable insights into the introduced RNN and CNN-based models. The motivation behind this choice is the need for solutions that provide an explanation based on the whole dataset, rather than being based on single samples. For this reason, other popular XAI techniques such as Local Interpretable Model-agnostic Explanations (LIME)~\cite{ribeiro2016should} are disregarded. Similarly, SHapley Addictive exPlanations (SHAP)~\cite{lundberg2017unified} is not contemplated, as it would not allow us to compare CNN results to RNN due to the difference in calculating their Shapley coefficients.

Model interpretation through SmoothGrad rests upon the definition of the sensitivity maps $M_i(\bm{x}(n))$, which aims to differentiate the prediction function $S_i(\bm{x}(n))$ for each output feature $i \in [1, \dots, q]$ with respect to the input $\bm{x}(n)$. The algorithm proposed in \cite{SmoothGrad} is adopted, in which the authors modified its standard calculation to compensate for the rapid fluctuations of the computed partial derivatives. The final sensitivity map results from the average of $N$ sensitivity maps $M_i(\bm{x}(n))$ computed on different perturbed versions of $\bm{x}(n)$ through Gaussian kernels, and in detail: 
\begin{equation}
 	\hat{M}_i(\bm{x}(n))=\frac{1}{N} \sum_{k=1}^{N} M_i\left(\bm{x}(n)+\epsilon_k\right)\text{ with }i \in [1, \dots, q].
 \end{equation}
where ($\epsilon_1, \dots, \epsilon_N$) are sampled from a Gaussian random variable $\mathcal{N}\left(0, \sigma^{2}\right)$, and $\sigma$ measures the intensity of the perturbation.

Instead, LRP defines the metric relevance, which is calculated in an iterative fashion from the output to the input neurons. The idea is to start with a neuron $k$ representing the output layer and calculate the relevance value of the neuron at the previous layer $j$ using this equation:

\begin{equation}
 	R_j = \sum_{k} \frac{a_j w_{jk}}{\sum_{0,j} a_j w_{jk}} R_k
 \end{equation}
where $a_j$ denotes the activation of the neuron $j$, and $w_{jk}$ is the weight between the two neurons $i$ and $j$. 
This formula is used iteratively to compute the relevance $R$ for every neuron of the previous layer. 
The process is then repeated layer by layer until reaching the input. The obtained final metric (from output to input layers) shows which input features are most relevant to the output. 

The two XAI algorithms have been implemented using the \texttt{iNNvestigate} Python library \cite{iNNvestigate2019}.

\subsection{Explainability Analysis}\label{subsec:explainability_results}

It is to be noted that both SmoothGrad and LRP are meant to be applied to a single specific input sample. Instead, the aim here is to interpret the behavior of the models on the whole dataset. Therefore, the whole $\mathcal D_t^{\text{PS}}$ dataset is used to obtain the correspondent sensitivity and relevance values. For the sake of visibility, the next figures show the squared average of the resulting values and highlight the input features that play a key role in the prediction across the studied dataset, independently of the sign of their contribution.
In line with the \mbox{XAI} paradigm, all the processed results have been plotted in groups of five heatmaps, each one corresponding to one of the $q$ output labels, i.e.,  $RNTI_{count}$, $RB_{down}$, $RB_{up}$, $THR_{down}$, and $THR_{up}$.
The \mbox{$x$-axis} contains the $m$ input variables, and the \mbox{$y$-axis} shows the progressive temporal index of the $p$ input observations.
Sensitivity and relevance are reported in the table with a color format. Darker shades represent features with higher values, i.e., more important in making predictions.

\begin{figure*}
    \centering
    \begin{subfigure}{.5\textwidth}
      \centering
      \includegraphics[width=.95\linewidth]{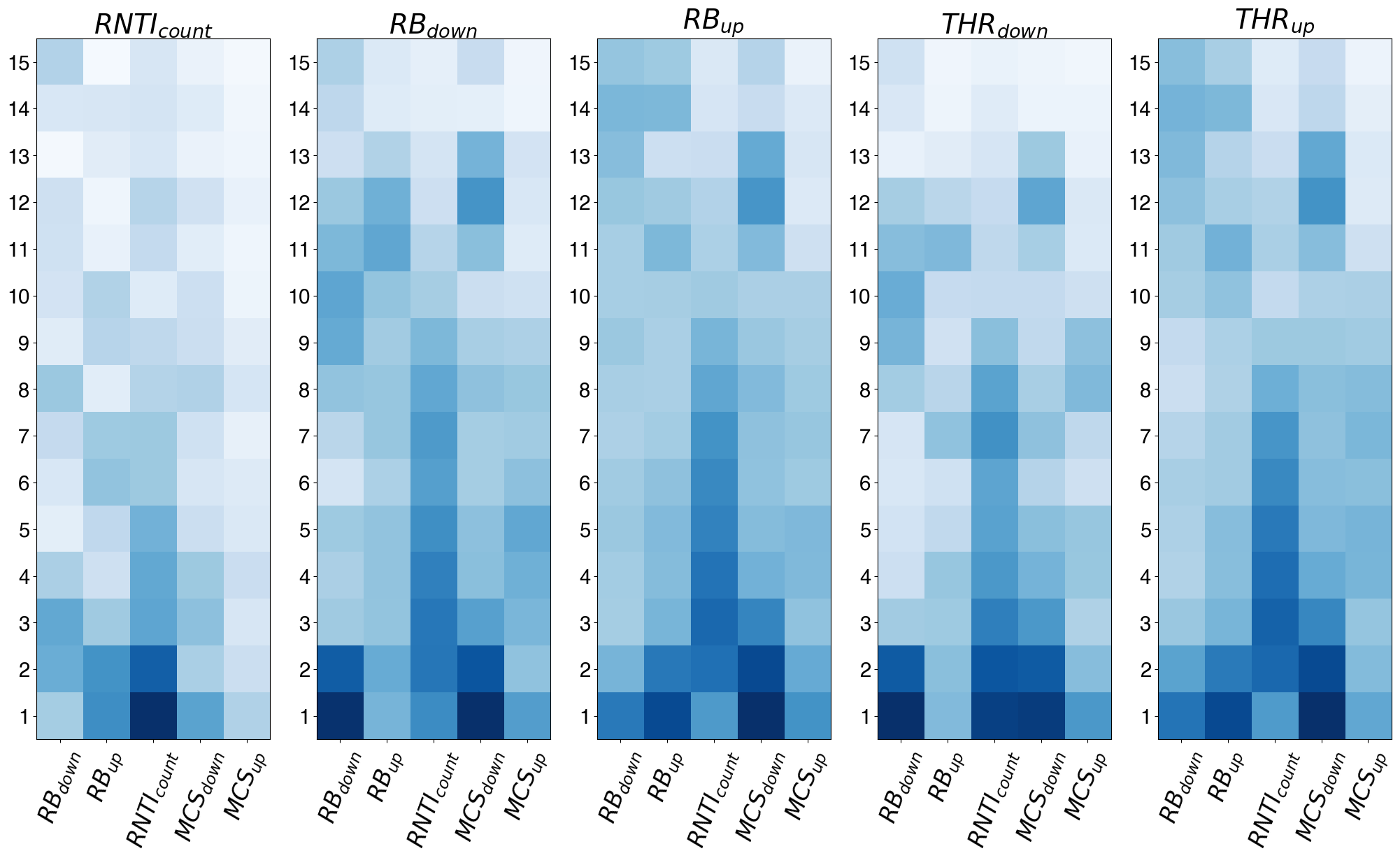}
      \caption{SmoothGrad}
      \label{fig:CNN_p15_GS0_SG}
    \end{subfigure}%
    \begin{subfigure}{.5\textwidth}
      \centering
      \includegraphics[width=.95\linewidth]{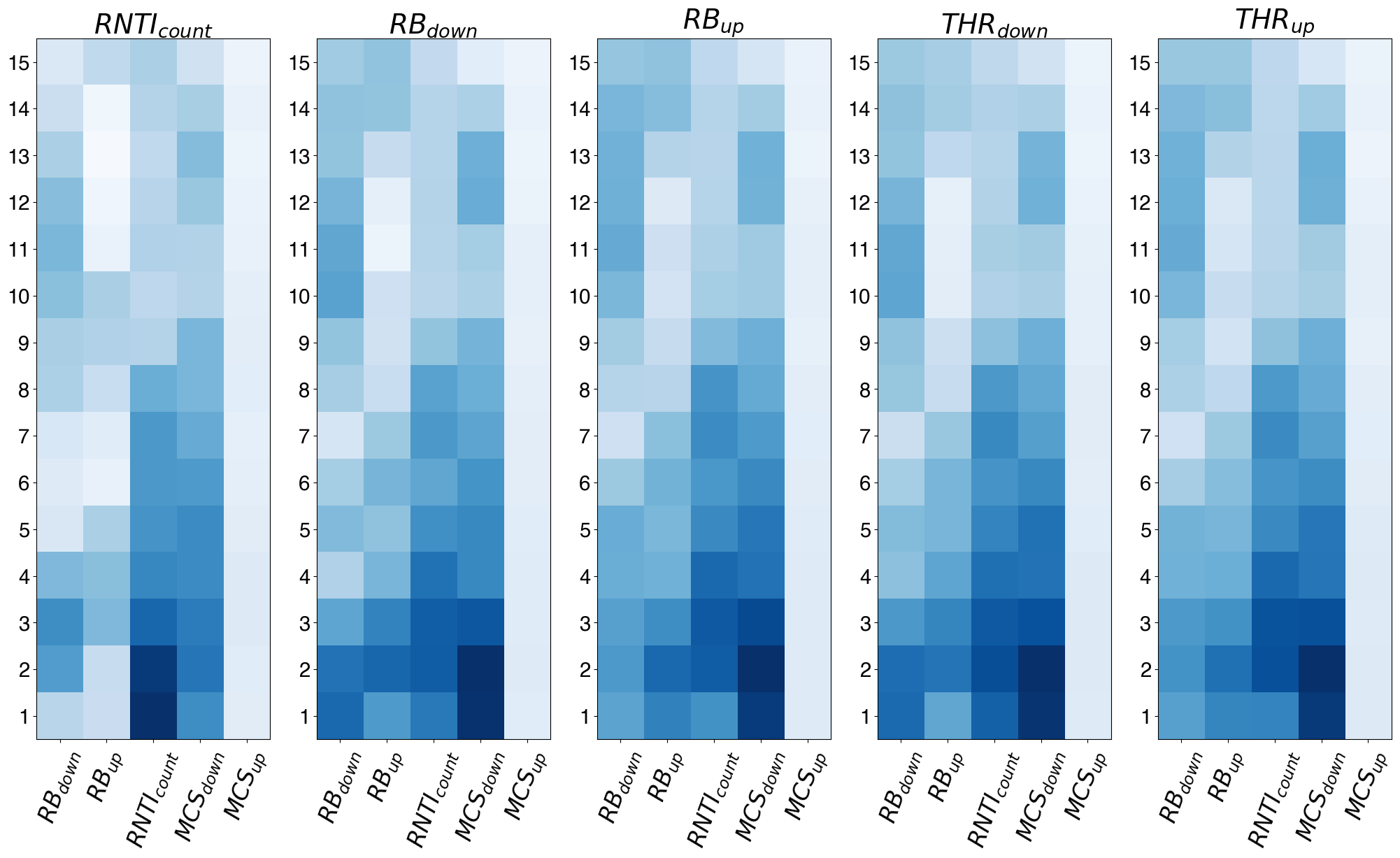}
      \caption{LRP}
      \label{fig:CNN_p15_GS0_LRP}
    \end{subfigure}
    \caption{SmoothGrad and LRP algorithms' heatmaps when applied to the CNN model, $p=15,$ $\Delta n = 0.$}
    \label{fig:CNN_p15_GS0}
\end{figure*}

\begin{figure*}
    \centering
    \begin{subfigure}{.5\textwidth}
      \centering
      \includegraphics[width=.95\linewidth]{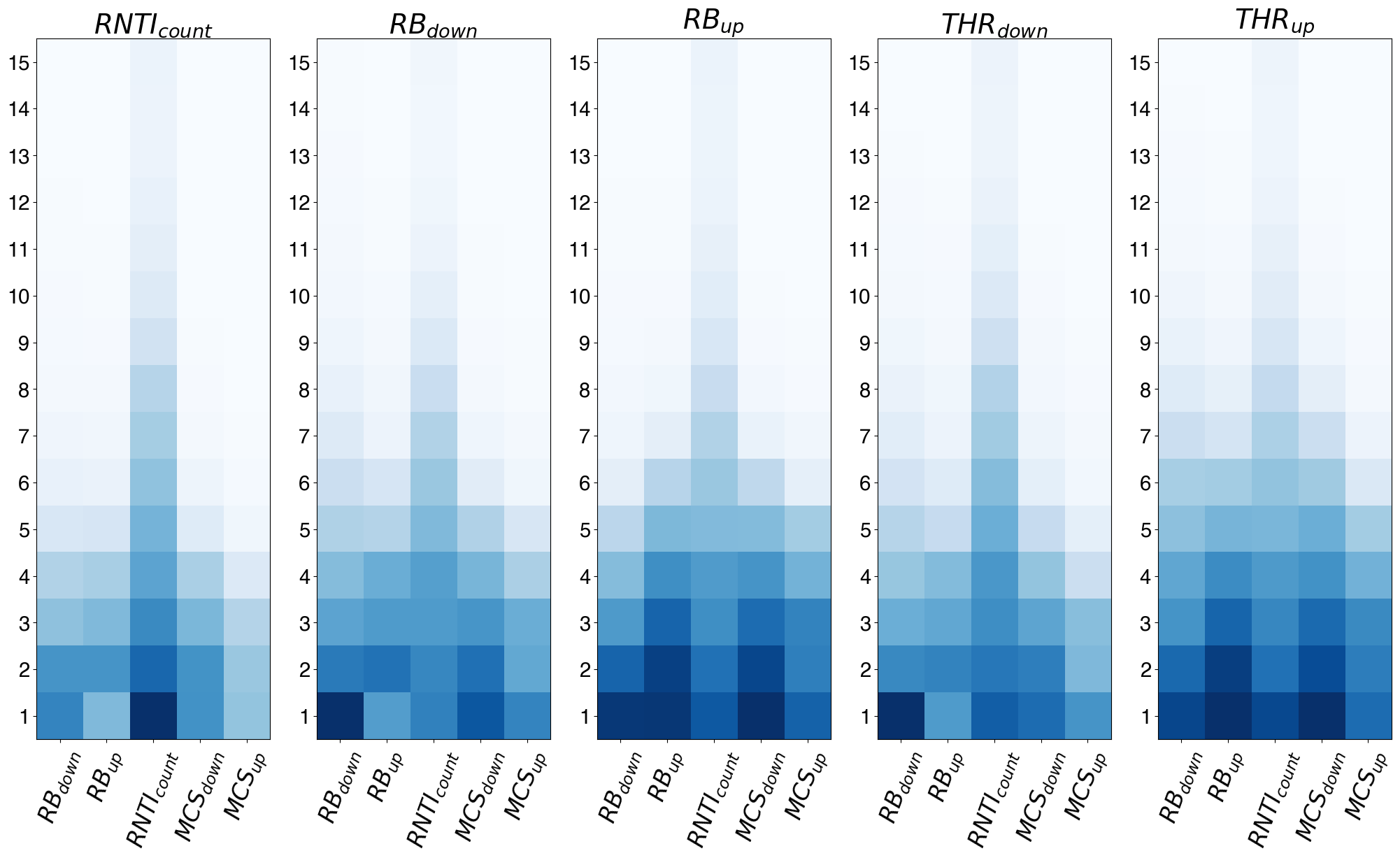}
      \caption{SmoothGrad}
      \label{fig:RNN_p15_GS0_SG}
    \end{subfigure}%
    \begin{subfigure}{.5\textwidth}
      \centering
      \includegraphics[width=.95\linewidth]{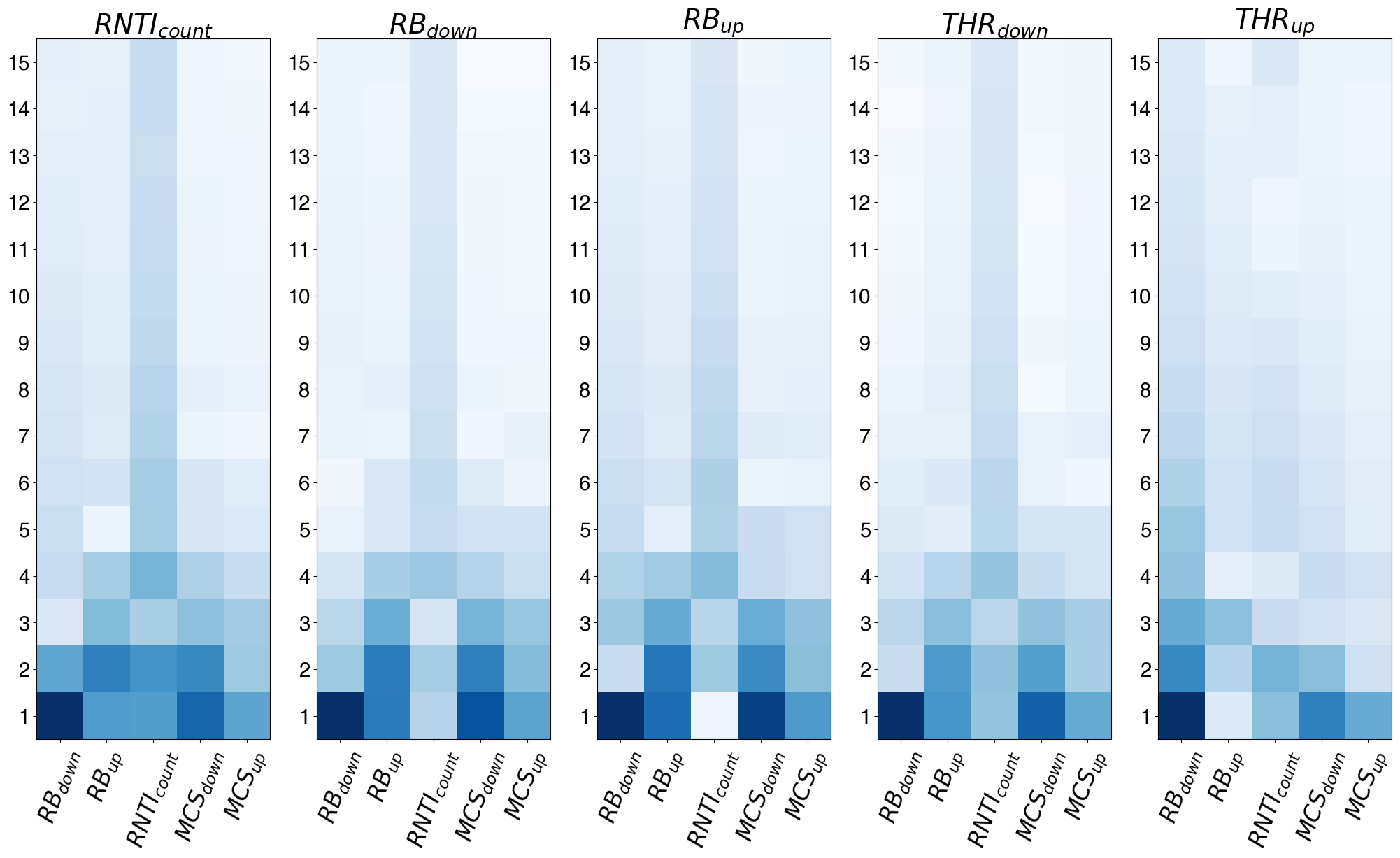}
      \caption{LRP}
      \label{fig:RNN_p15_GS0_LRP}
    \end{subfigure}
    \caption{SmoothGrad and LRP algorithms' heatmaps when applied to the RNN model, $p=15,$ $\Delta n = 0.$}
    \label{fig:RNN_p15_GS0}
\end{figure*}

\begin{figure*}
    \centering
    \begin{subfigure}{.5\textwidth}
      \centering
      \includegraphics[width=.95\linewidth]{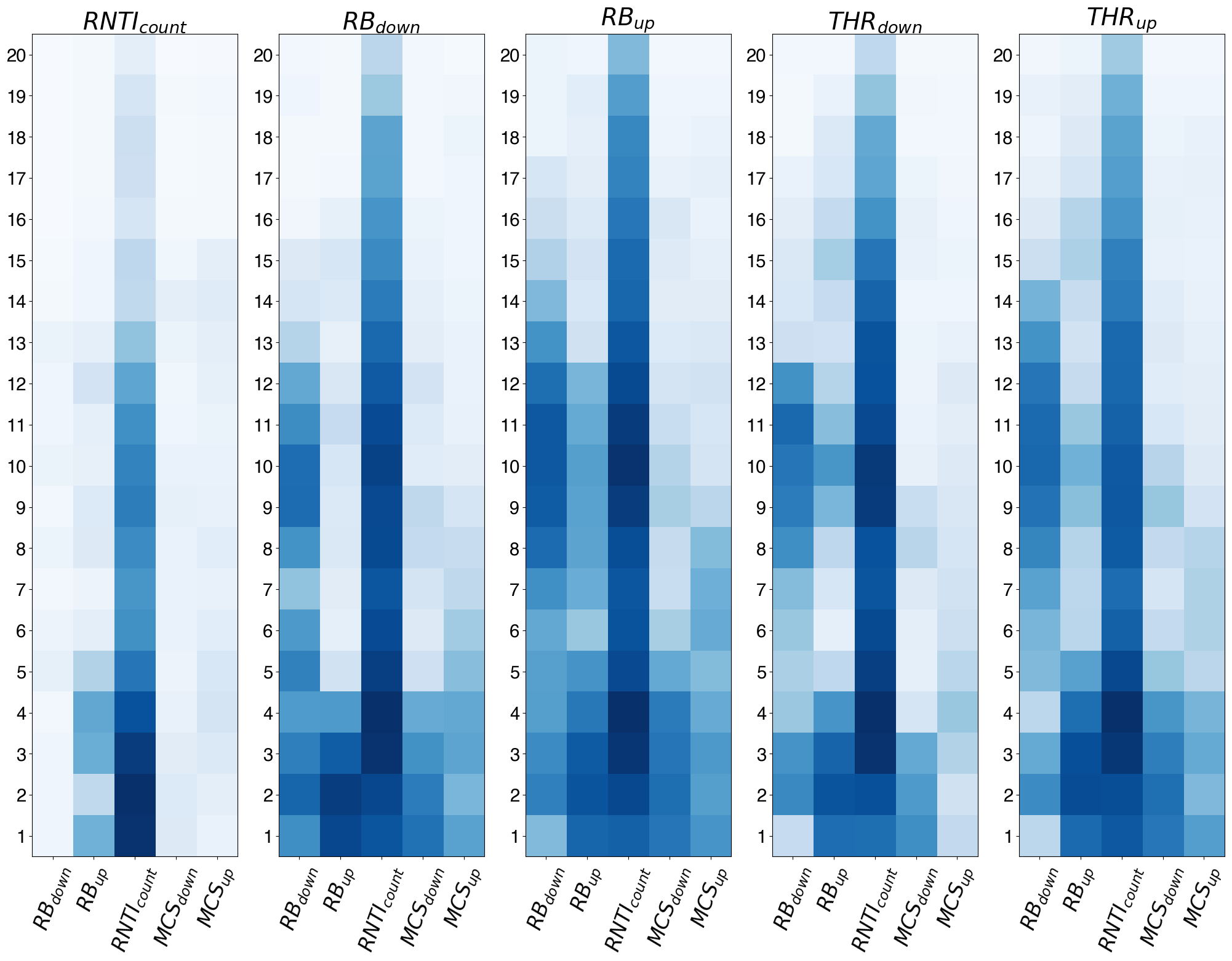}
      \caption{SmoothGrad}
      \label{fig:CNN_p20_GS14_SG}
    \end{subfigure}%
    \begin{subfigure}{.5\textwidth}
      \centering
      \includegraphics[width=.95\linewidth]{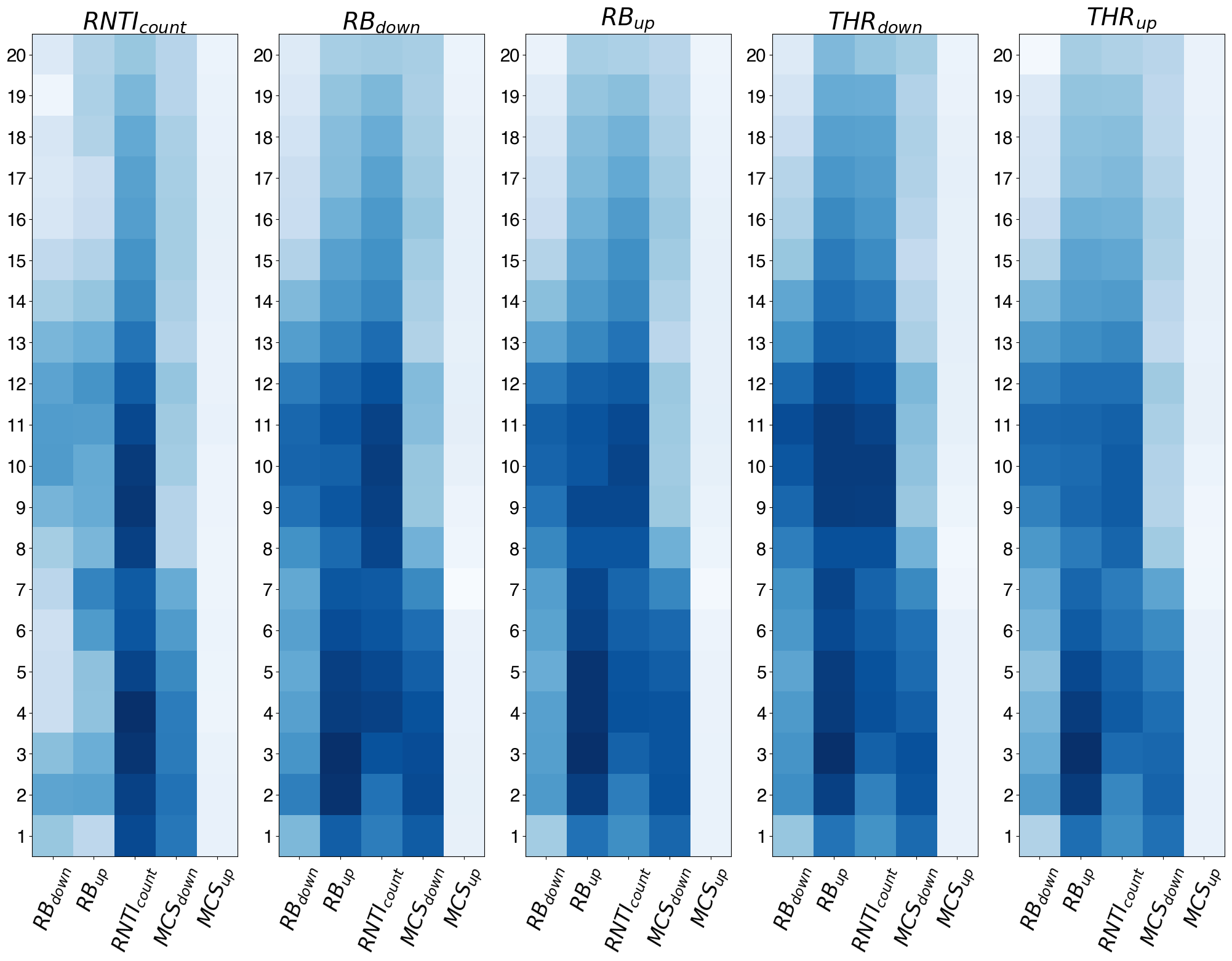}
      \caption{LRP}
      \label{fig:CNN_p20_GS14_LRP}
    \end{subfigure}
    \caption{SmoothGrad and LRP algorithms' heatmaps when applied to the CNN model, $p=20,$ $\Delta n = 14.$}
    \label{fig:CNN_p20_GS14}
\end{figure*}

\begin{figure*}
    \centering
    \begin{subfigure}{.5\textwidth}
      \centering
      \includegraphics[width=.95\linewidth]{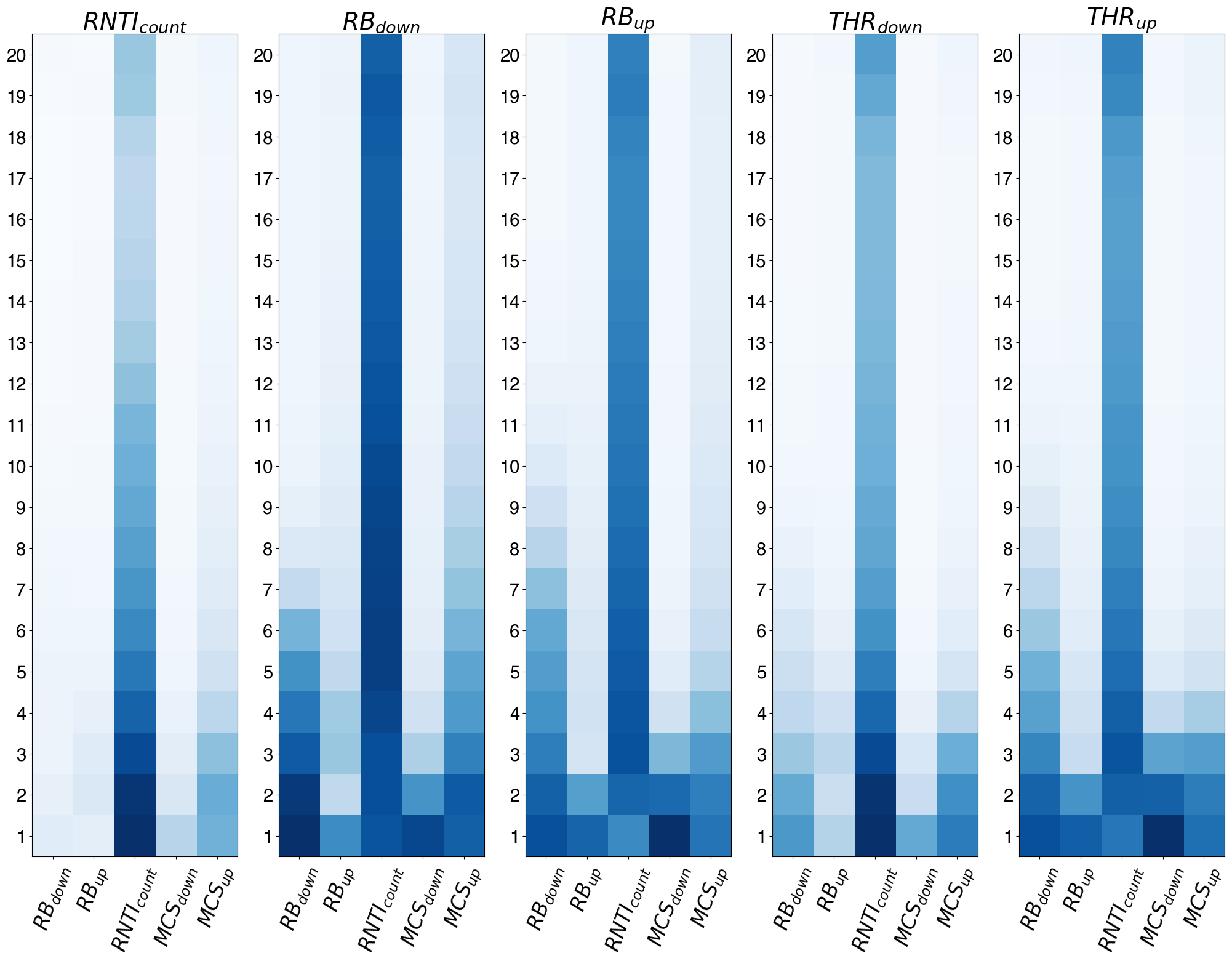}
      \caption{SmoothGrad}
      \label{fig:RNN_p20_GS14_SG}
    \end{subfigure}%
    \begin{subfigure}{.5\textwidth}
      \centering
      \includegraphics[width=.95\linewidth]{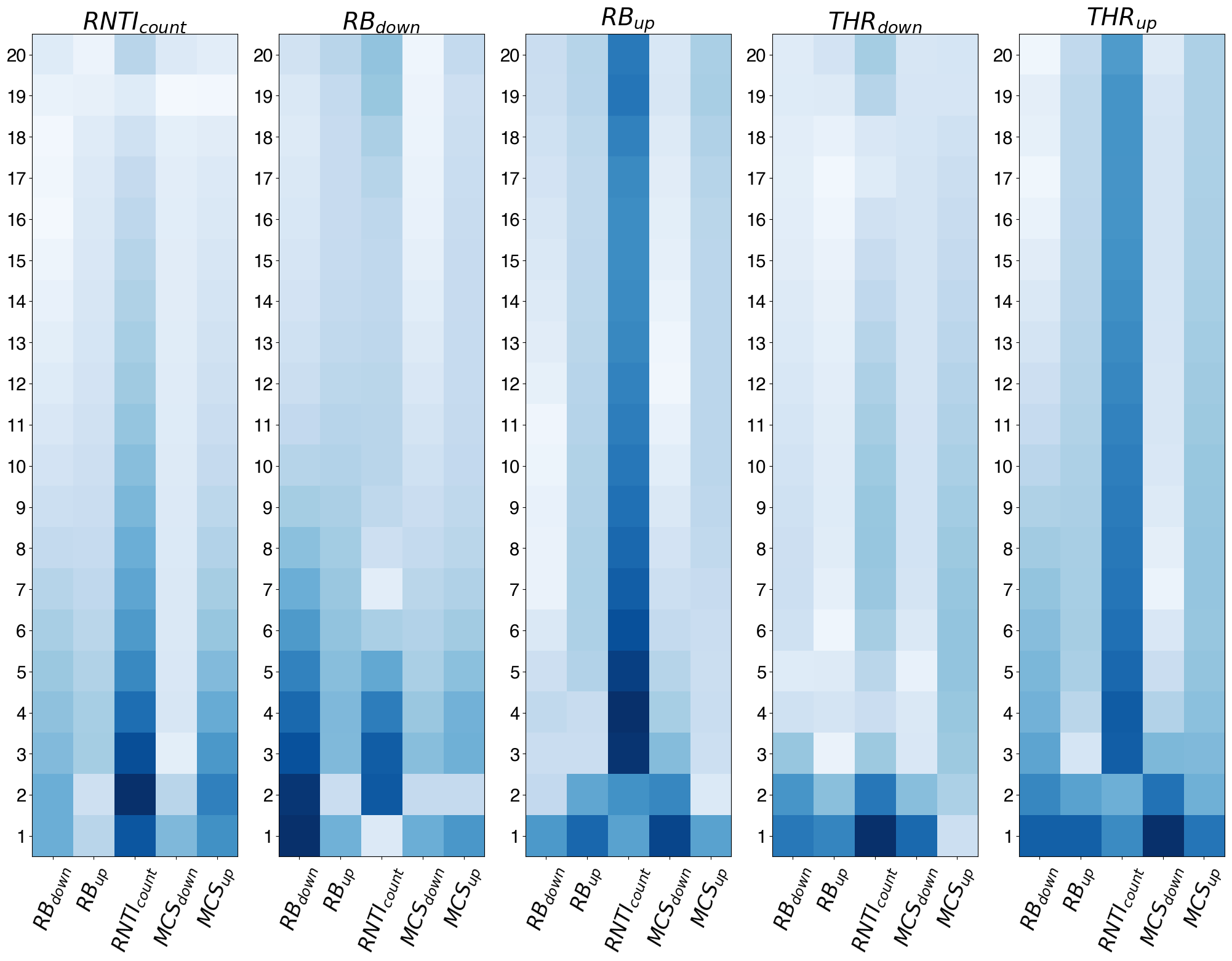}
      \caption{LRP}
      \label{fig:RNN_p20_GS14_LRP}
    \end{subfigure}
    \caption{SmoothGrad and LRP algorithms' heatmaps when applied to the RNN model, $p=20,$ $\Delta n = 14.$}
    \label{fig:RNN_p20_GS14}
\end{figure*}

Fig.~\ref{fig:CNN_p15_GS0_SG} and Fig.~\ref{fig:CNN_p15_GS0_LRP} contain the visual representation of the output heatmaps of the SmoothGrad and LRP algorithms, respectively, applied to the CNN model trained on $\mathcal D_t^{\text{PS}}$ with a BS of $256$ and with $\Delta n = 0$. 
Fig.~\ref{fig:CNN_p15_GS0_SG} indicates that the \textit{RNTI}$_{count}$ feature is the most informative for predicting \textit{RNTI}$_{count}$. Instead, \textit{RB}$_{down}$ and \textit{THR}$_{down}$ predictions depend strongly on \textit{RB}$_{down}$ and \textit{MCS}$_{down}$; \textit{RB}$_{up}$ and \textit{THR}$_{up}$ predictions are mainly influenced by \textit{MCS}$_{down}$, \textit{RB}$_{up}$ and \textit{RNTI}$_{count}$. 
This means that forecasting \textit{RNTI}$_{count}$ mainly depends on the historical data of \textit{RNTI}$_{count}$ itself,  which is due to the fact that the number of users is the variable with the lowest variance in time, as well as to the small $\Delta n$ value.
In particular, the low variability allows for accurate predictions using the last input observations.

The LRP heatmap in Figure~\ref{fig:CNN_p15_GS0_LRP} presents similar results to SmoothGrad and confirms the dependencies among input features and predicted labels.
Again, it confirms that \textit{RNTI}$_{count}$ predictions strongly rely on the historical data of the same variable, as noticed with SmoothGrad. 
Moreover, \textit{MCS}$_{up}$ appears to be almost irrelevant for the predictions (even less than with SmoothGrad), probably due to the limited traffic in uplink from the captured datasets.
The entries in the upper part of the heatmaps (i.e., higher $p$) result to be less significant.
The results in Fig.~\ref{fig:CNN_p15_GS0_SG} and Fig.~\ref{fig:CNN_p15_GS0_LRP} confirm the analysis in Section~\ref{subsec:results_FS}, where we discussed that longer observation windows do not increase the prediction accuracy.
On the other hand, it emerges that \textit{MCS}$_{down}$ plays an important role, likely due to its central position in the input tensor and its correlation with the variables lying on the borders, i.e., \textit{RB}$_{down}$ and \textit{MCS}$_{up}$.

Figure~\ref{fig:RNN_p15_GS0_SG} and Figure~\ref{fig:RNN_p15_GS0_LRP} contain the heatmaps of SmoothGrad and LRP applied to the RNN model trained on $\mathcal D_t^{\text{PS}}$ with a BS of $128$ and for $\Delta n=0$. Results indicate a different relationship between input features and predicted labels compared to the CNN model. Figure~\ref{fig:RNN_p15_GS0_SG} confirms that the \textit{RNTI}$_{count}$ feature is the most informative for predicting \textit{RNTI}$_{count}$. Instead, \textit{RB}$_{down}$ and \textit{THR}$_{down}$ predictions depend strongly on \textit{RB}$_{down}$ and \textit{MCS}$_{down}$ and \textit{RNTI}$_{count}$; \textit{RB}$_{up}$ and \textit{THR}$_{up}$ predictions are instead influenced by all the input variables. Fig.~\ref{fig:RNN_p15_GS0_LRP} shows that \textit{RNTI}$_{count}$ predictions depend mainly on \textit{RB}$_{down}$. 
No other significant differences can be appreciated with respect to Figure~\ref{fig:RNN_p15_GS0_SG}.
Both figures highlight that longer input observations do not have any influence on the predicted labels for RNN models, with the only exception in the \textit{RNTI}$_{count}$ case.

In Figure~\ref{fig:CNN_p20_GS14} and Figure~\ref{fig:RNN_p20_GS14}, the analysis on the future predictions $\Delta n$ is presented, by showing the heatmaps of the two XAI algorithms with CNN and RNN models trained on $\mathcal D_t^{\text{PS}}$ with $\Delta n=14$. In Section~\ref{subsec:results_FS}, we discussed that the CNN model generally performs better in tasks where $\Delta n$ is large, indicating that CNNs can extract much more information from the input samples compared to RNNs. This is clearly visible in Figure~\ref{fig:CNN_p20_GS14} graphs, which show darker pixels for higher values of $p$ ($y$ axis) for both SmoothGrad and LRP algorithm.

\subsection{Future Research Directions}

Based on the outcomes of XAI analysis, several optimizations can be performed to the DL architectures used here for mobile traffic prediction.
Considering RNNs, a straightforward optimization is represented by the reduction of the input observation window ($p$) for the simplest tasks (i.e., for small $\Delta n$), which could lead to lower computational costs without jeopardizing the performance. Regarding more complicated extensions, the combination of multi-task \cite{Zhang2022_MultiTask} and ensemble model approaches \cite{ensemble_review_2016} could be considered for creating a meta-model where each output parameter has only the most important input features according to XAI outcomes. For example, $RB_{up}$ could be removed from the base estimators of the global ensemble model corresponding to several predictions where the importance of such an input feature has been shown as negligible by XAI. Similarly, $RNTI_{count}$ can be predicted almost entirely using previous $RNTI_{count}$ values, so simpler models can be used for this task.

On the other hand, XAI algorithms fail to provide a clear interpretation of the neurons' activation and interconnections. We believe this is a key open issue to enhance the black box approach to neural networks. With more information about the internal structure of the model, DTL layer freezing can be optimized towards a better energy-accuracy trade-off. 
In particular, LRP is suitable by design to extract networks' internal layers information, but the enormous number of neurons complicates the extraction of interpretable observations. To overcome this issue, the general global network can be divided into sub-networks that can be more easily managed with aggregated metrics. 
This would allow to evaluate relevance on a more \mbox{fine-grained} basis and could be used to detect insights about those sub-networks. This procedure has to be properly designed to find the correct dimension of the sub-network and, thus, avoid the introduction of more computational complexity compared to the one that would be saved by the DTL algorithm.
\section{Conclusions}
\label{sec:conclusion}

This paper presented a novel edge computing framework for mobile traffic prediction, utilizing three datasets obtained from decoding the unencrypted LTE control channel at the network edge.
Two main DL architectures were designed, based on CNNs and RNNs, using both a stand-alone approach and DTL techniques to reduce the required computational resources.

Results show that the CNN architectures outperform the RNNs in terms of accuracy, and in both cases DTL enables an accuracy improvement in 85\% of the studied cases with respect to the \mbox{stand-alone} counterpart.
Moreover, DTL is able to significantly decrease the computation complexity and thus the energy consumption relative to training, allowing an energy footprint reduction of 60\% and 90\% for CNNs and RNNs, respectively.
Finally, two \mbox{cutting-edge} XAI techniques were employed to explain DL models' prediction behavior and provide insights into the accuracy results.

The proposed approach faces challenges with Stochastic Gradient Descent optimization, particularly due to variations in the granularity and nature of the diverse inputs and outputs, as well as the significant presence of outliers in the data. This can impact the model's overall performance. To address this limitation, future work could consider adopting multi-task learning or ensemble model approaches to significantly enhance prediction accuracy and reduce energy consumption, offering a more efficient solution.

Based on the XAI outcomes, some possible extensions to this work have been derived. Regarding the architecture, the combination of multi-task and ensemble model approaches can be used for implementing a meta-model in which the prediction of each output parameter uses only its most important input features. For XAI, DTL layer freezing can be optimized by studying extensions of LRP that provide insights into the activation of internal neurons and interpretation of their interconnections. To achieve this, the general global network can be divided into sub-networks that are more easily managed with aggregated metrics.

{
    \small
    \bibliographystyle{ieeetr}
    \bibliography{biblio}
}

\begin{IEEEbiography}[{\includegraphics[width=1in,height=1.25in,clip,keepaspectratio]{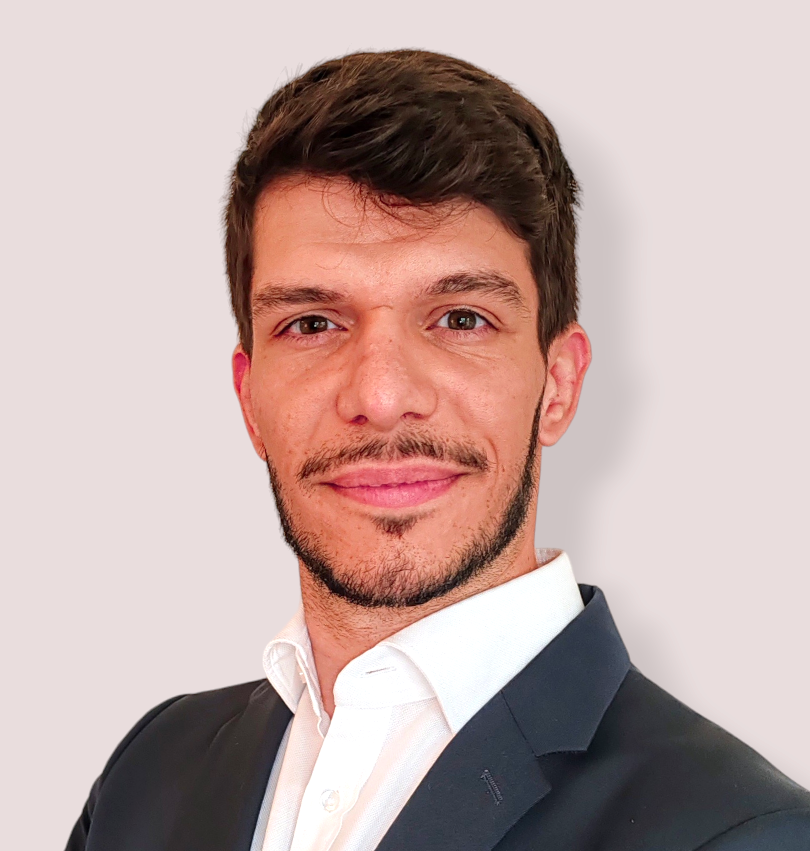}}]{Alfredo Petrella} received the B.Sc. degree in mathematics and the M.Sc. degree in data science from the University of Padua, Italy, in 2018 and 2021, respectively.
In 2021, he was a Research Assistant in the Sustainable Artificial Intelligence research unit at CTTC. Since 2022, he has been working as a consultant in Artificial Intelligence and Machine Learning in Milan, Italy, being deeply involved in both the theoretical and practical aspects of AI, and applying cutting-edge research to real-world problems.
His research interests include optimization models, deep learning, and natural language processing.
\end{IEEEbiography}

\begin{IEEEbiography}[{\includegraphics[width=1in,height=1.25in,clip,keepaspectratio]{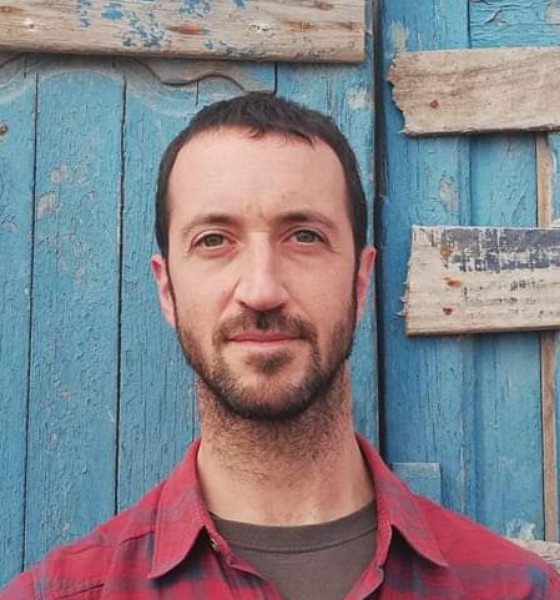}}]{Marco Miozzo}
received his M.Sc. degree in Telecommunication Engineering from the University of Ferrara (Italy) in 2005 and the Ph.D. from the Technical University of Catalonia (UPC) in 2018. In June 2008 he joined the Centre Tecnologic de Telecomunicacions de Catalunya (CTTC).
His main research interests are: sustainable mobile networks, green wireless networking, multi-agent systems, machine learning, green AI, distributed and collaborative learning, and pervasive AI.
\end{IEEEbiography}

\begin{IEEEbiography}[{\includegraphics[width=1.1in,height=1.3in,clip,keepaspectratio]{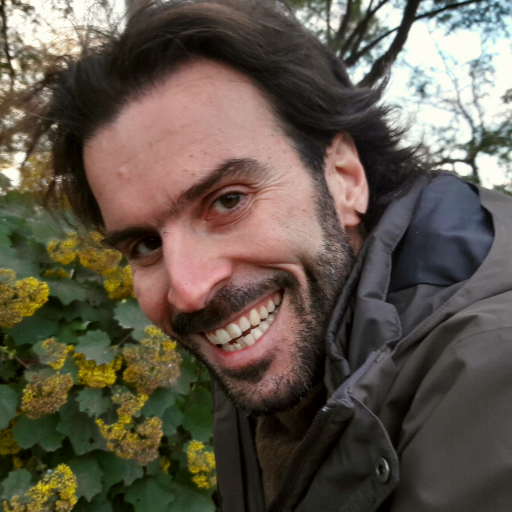}}]{Paolo Dini}
received M.Sc. and Ph.D. from the University of Roma La Sapienza, in 2001 and 2005, respectively. He is currently a Leading Researcher with the Centre Tecnologic de Telecomunicacions de Catalunya (CTTC) and coordinates the activities of the Sustainable AI research unit. His current research interests include sustainable computing for cyber-physical systems, distributed optimization and optimal control, machine learning, multi-agent systems.
He received two awards from the Cisco Silicon Valley Foundation for his research on distributed control over heterogeneous mobile
networks, in 2008 and 2011, respectively. 
He has been involved in more than 20 research and development projects. He is currently the Coordinator of CHIST-ERA SONATA project on sustainable computing and communication at the edge and the Scientific Coordinator of the EU H2020 MSCA Greenedge
European Training Network on edge intelligence and sustainable computing.
\end{IEEEbiography}

\EOD

\end{document}